\documentclass[aps,pra,showpacs,twoside,twocolumn,10pt]{revtex4-1}
\usepackage{newlfont}
\usepackage{color}  
\usepackage[colorlinks=true, citecolor=red, urlcolor=blue ]{hyperref} 
\usepackage{amssymb}
\usepackage{amsfonts}
\usepackage{amsmath}
\usepackage{wasysym}
\usepackage{graphicx}
\usepackage{epsfig}
\usepackage{amsthm}
\usepackage{bm}
\usepackage{multirow}
\usepackage{ulem}
\usepackage{mathrsfs}
\usepackage{epstopdf}
\usepackage{theorem}
\definecolor{indiagreen}{rgb}{0.07, 0.53, 0.03}

\usepackage{subfigure,palatino,mathtools,braket,times,soul}

\begin{document}

\title{
Quantum Dense Coding Network using Multimode Squeezed States of Light }

\author{Ayan Patra\(^1\), Rivu Gupta\(^1\), Saptarshi Roy\(^{2,3}\), Tamoghna Das\(^4\), Aditi Sen(De)\(^1\)}

\affiliation{\(^1\)Harish-Chandra Research Institute,  A CI of Homi Bhabha National
Institute, Chhatnag Road, Jhunsi, Prayagraj - 211019, India}
\affiliation{\(^2\) Quantum Information and Computation Initiative, Department of Computer Science, The University of Hong Kong, Pokfulam Road, Hong Kong.}
\affiliation{\(^3\) HKU-Oxford Joint Laboratory for Quantum Information and Computation.}
	\affiliation{\(^4\)International Centre for Theory of Quantum Technologies,  University of Gda\'{n}sk, 80-952 Gda\'{n}sk, Poland.}

\begin{abstract}

We present a framework of a multimode dense coding network with multiple senders and a single receiver using continuous variable systems. The protocol is scalable to arbitrary numbers of modes with the encoding being displacements while the decoding involves homodyne measurements of the modes after they are combined in a pairwise manner by a sequence of beam splitters, thereby exhibiting its potentiality to implement in laboratories with currently available resources. We compute the closed-form expression of the dense coding capacity for the cases of two and three senders that involve sharing of three- and four-mode states respectively. The dense coding capacity is computed with the constraint of fixed average energy transmission when the modes of the sender are transferred to the receiver after the encoding operation. In both cases, we demonstrate the quantum advantage of the protocol using paradigmatic classes of three- and four-mode states. The quantum advantage increases with the increase in the amount of energy that is allowed to be transmitted from the senders to the receiver.


\end{abstract}

\maketitle

\section{Introduction}

Non-classical correlations  play a crucial role in building quantum information technologies like  quantum cryptography \cite{BB84, EkertCrypto, Crypto3, Crypto4,Vazirani_PRL_2014,Mayers_arxiv_1998,Miller_JACM_2016}, dense coding (DC) \cite{bennettwiesner, AditiComm, GisinComm, AditiComm2, AditiComm3}, teleportation \cite{BBCJPW,Vaidman94,Braunstein_PRL_1998}, one-way quantum computation \cite{Raussendorf_PRL_2001,Briegel_PRL_2001,Raussendorf_PRA_2003,Walther_Nature_2005,Raussendorf_NJP_2007,Raussendorf_PRL_2007,Verstraete_Nature_2009}, and random number generation \cite{Ma_NPJ_2016,Kollmitzer_Springer_2020} to name a few.  Among them, the dense coding protocol is essential for  transmitting classical information without security from one place to another with the help of a shared entangled state, which exhibits improvements in capacity over its classical counterparts.  
The original DC proposal with point-to-point communication was later extended to multiparty networks involving multiple senders and a single as well as two receivers \cite{Bruss,DCCamader,DCCTamo1,Tamoghna,Prabhu} although such a design of networks is mostly limited to the finite-dimensional systems (cf. \cite{Lee_PRA_2014,Czekaj_PRA_2010}). Interestingly, it was shown that even in the case of quantum key distribution, it is beneficial to apply the secure dense coding protocol, as it doubles  the rate of secure key per transmitted qubit  between the honest parties, and also increases the chance of detecting  the presence of malicious eavesdropper, up to two senders in the single receiver scenario \cite{Bostrom_PRL_2002,Beaudry_PRA_2013,Das_arXiv_2021}. 

Continuous variable (CV) systems provide an important platform for realizing quantum protocols. It can overcome several limitations arising in the finite-dimensional case, 
a prominent one being the  distinction of four orthogonal Bell states with linear optical elements required in the stage of decoding of classical information \cite{Weinfurter_1994, DC_obstacle, Lutkenhaus_PRA_1999,DC_obstacle2, DC_obstacle3}.
However, these drawbacks can be overcome when one considers the  continuous variable (CV) systems, in particular, the mode-entanglement of  multi-photon quantum optical systems, where the average number of photons in a mode is taken to be arbitrary.  
The pioneering work on dense coding in the field of CV systems (which we refer to as CVDC)  was first proposed by Braunstein and Kimble \cite{Braunstein_PRA_2000} where the Einstein Podolsky Rosen (EPR)  state \cite{Einstein_PRA_1935} is shared between a single sender and a single receiver to transfer classical information. The encoding operation is performed  by applying the displacement operator, which is distributed according to a Gaussian distribution of vanishing mean and variance $\sigma$.
In recent years, lots of developments have been made for the successful realization of classical information transmission in CV systems \cite{Hao_PRL_2021,Barzanjeh_PRA_2013}, particularly with shared Gaussian entangled  states between the sender and the receiver \cite{Adesso_OSID_2014,Kim_PRA_2002,Wang_OptEx_2020,Ralph_PRA_2002,Jing_PRL_2003}. 

In this paper,  we design a framework for the dense coding protocol involving an arbitrary number of senders and a single receiver  with quantum optical fields.
Each of the senders performs local unitary encoding with the help of the displacement operator, drawn uniformly from a Gaussian distribution with variance $\sigma$. Thereafter, the modes are transmitted to the receiver, who combines the modes pairwise with the help of the beam splitters for decoding the message sent by the sender. The transmission coefficients of the beam splitters have been kept arbitrary, so as to determine the decoding configuration which can overcome the classical bound. The proposed procedure works with an arbitrary number of senders and a single receiver. 

 When two and three senders  share three- and four-mode genuinely entangled Gaussian states with a single receiver respectively, we exhibit quantum advantage, i.e., when the capacity of the quantum protocol beats the classical threshold value for a given energy,  which can be obtained between the arbitrary number of senders and a receiver without any shared entanglement. Specifically, 
we identify the region characterized by the state parameters which lead to  quantum advantage.
The DC protocol in CV systems, required to obtain a valid classical capacity, is typically implemented with a fixed average number of photons in the sender modes which bounds the energy of the system. We report here the threshold photon number necessary for outperforming the classical routine. Moreover, the initial squeezing strength leading to a quantum benefit is determined for some classes of paradigmatic three- and four-mode states which manifest that the current state-of-the-art experiments can achieve quantum advantage in a DC network.

The paper is organized in the following way. The multimode dense coding scenario is introduced in Sec. \ref{sec:pre} which includes encoding and decoding of classical information,   the capacity of DC via quantum protocol, and the corresponding classical scheme without the shared entangled state. We then illustrate the DC capacities when the senders and a receiver share three- and four-mode channels in Secs. \ref{sec:3modecap} and \ref{sec:four-mode-dc} respectively. The comparisons of DC capacities with classical protocols are also discussed in these sections while concluding remarks are in Sec. \ref{sec:conclu}. 

\begin{figure}
    \centering
    \includegraphics[width=\linewidth]{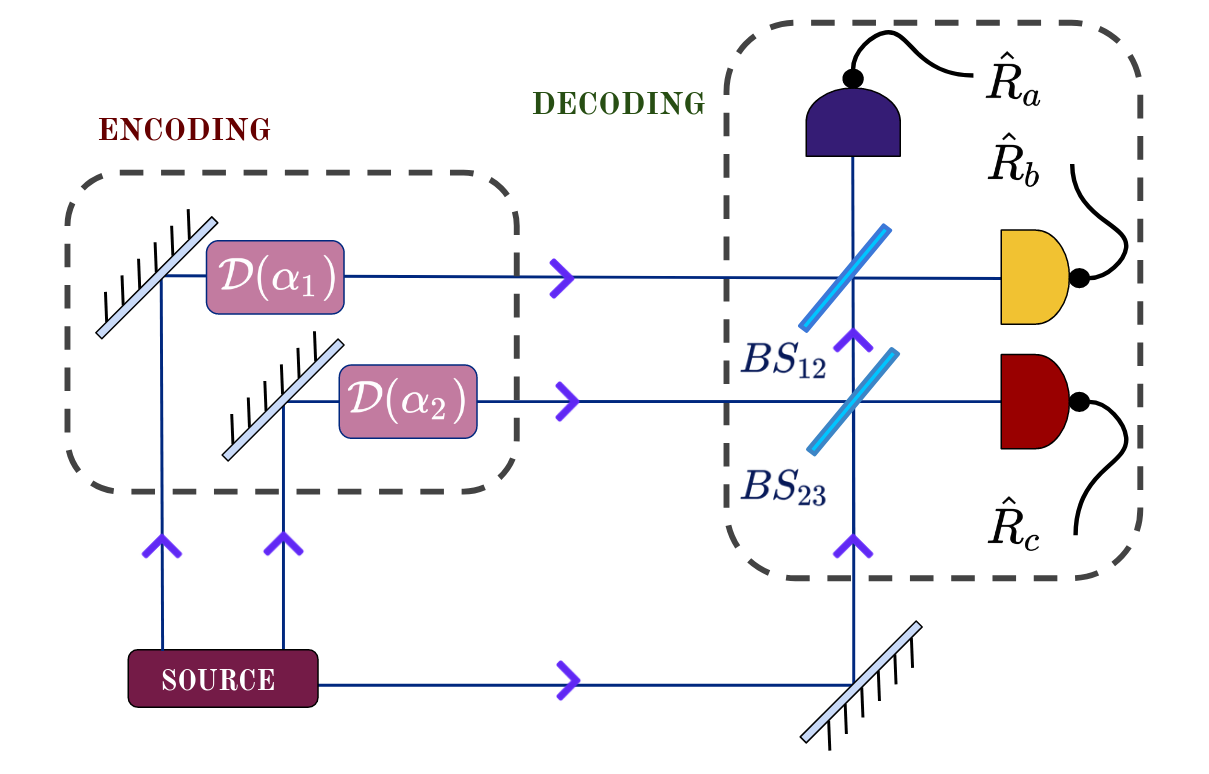}
    \caption{Schematic diagram for the CVDC protocol involving two senders and a single receiver. The schematic has two components in DC -- encoding at the senders' ends which involve the displacement operators, denoted by \(D(\alpha_k)\) \((k =1, 2)\), and the decoding part in the receiver's side after obtaining modes from the sender upon encoding. The latter one requires a combination of  beam splitters, \(BS_{12}\), and  \(BS_{23}\) and  homodyne measurements of quadratures, $\hat{R}_a$, $\hat{R}_b$, and $\hat{R}_c$ at the receiver's end.} 
    \label{fig:schematics}
\end{figure}

\section{Framework for Multimode Dense coding network}
\label{sec:pre}

We now introduce  
the formalism of the multimode dense coding network involving multiple senders and a single receiver necessary for our investigation  (see Fig. \ref{fig:schematics} for the case of two senders and a single receiver). We start by briefly recapitulating the basic properties of Gaussian states, and describe how they can be characterized by their first two moments in the phase space formalism. We also elucidate the Wigner function formalism, which turns out to be useful in the study of DC in continuous variable systems, and present the dense coding routine for classical information transfer between multiple senders and a single receiver. We focus on the multimode entangled states which are necessary for the successful implementation of the process and move on to construct the encoding and decoding schemes to arrive at an expression for the multimode dense coding capacity. Finally, we derive the classical capacity for multi-sender dense coding using continuous variable states without entanglement, which sets a benchmark on the classical bound for accessing the quantum advantage of the protocol.

\subsection{Multimode Gaussian states as resources}

Gaussian states are completely characterized by their displacement vector $\textbf{d}$ and covariance matrix $\Xi$ \cite{Adesso_OSID_2014}, given by
\begin{equation}
d_i = \langle \hat{R}_i \rangle,
\end{equation}
and
\begin{equation}
\Xi_{ij} = \frac{1}{2}\langle \hat{R}_i\hat{R}_j + \hat{R}_j \hat{R}_i\rangle - \langle \hat{R}_{i} \rangle \langle \hat{R}_{j} \rangle,
\end{equation}
where $\hat{R}_i$s are the phase space quadrature operators, $\hat{\textbf{R}} = (\hat{q}_1,\hat{p}_1,....,\hat{q}_\mathcal{N},\hat{p}_\mathcal{N})^T$,  satisfying the canonical commutation relation (CCR), $[\hat{R}_i,\hat{R}_j] = iJ_{kl}$. Here  $J$ is the $\mathcal{N}$-mode symplectic form,
$ J = \bigoplus\limits_{i=1}^{\mathcal{N}} \Omega $, where 
\[\Omega=
\begin{bmatrix}
0 & 1\\
-1 & 0
\end{bmatrix}.
\]
Therefore,  the transformations which preserve the CCR are symplectic, i.e., $SJS^T = J$.

In the phase space formalism of CV systems, the states can  equivalently be characterized by the characteristic function  \cite{Barnett_book_2002}
which reads, for an $\mathcal{N}$-mode state $\rho$,  
 as
\begin{eqnarray}
\chi_\rho (\bm{\alpha}) = \text{Tr} [\rho \hat{D}(\bm{\alpha})],
\end{eqnarray}
where $\bm{\alpha} = (\alpha_1, \alpha_2, \ldots \alpha_\mathcal{N})$ and $\hat{D}(\bm{\alpha}) = \bigotimes_{i = 1}^\mathcal{N} \hat{D}(\alpha_i)$ with $\hat{D}(\alpha_k) = \exp(\alpha_k \hat{a_k}^\dagger - \alpha_k^*\hat{a_k})$ being the displacement operator for mode $k$. The Fourier transform of the characteristic function is the well-known Wigner function \cite{Wigner_PR_1932}, which for a $\mathcal{N}$-mode Gaussian state, turns out to be a $2\mathcal{N}$-variable Gaussian function, given by \cite{Adesso_OSID_2014}
\begin{equation}
W(\textbf{R}) = \dfrac{\exp[-\frac{1}{2}(\textbf{R} - \textbf{d})^T \Xi^{-1} (\textbf{R} - \textbf{d})]}{(2\pi)^\mathcal{N}\sqrt{\det(\Xi)}}.
\end{equation}
Operationally, the reduced Wigner function obtained by integrating over the quadrature variables of $m$-modes gives the marginal probability distribution for the rest of the modes.



\subsection{Elements of CV Dense Coding with multiple senders}
\label{subsec:cv_multi}


Let us present here important constituents of the DC network with CV systems. 
One of the main ingredients of the prescribed  protocol involving multiple senders and a single receiver is the  class of $(\mathcal{N} - 1)$-parameter family of $\mathcal{N}$-mode Gaussian states shared between $(\mathcal{N} - 1)$ senders and a single  receiver. To implement  successful DC, we require suitable encoding of classical information by the senders and the corresponding decoding procedure by the receiver after all the modes have been transferred to the receiver. The success of  the protocol can be measured by computing  the multimode dense coding capacity. The quantum advantage of the protocol can only be guaranteed when the DC capacity crosses the  classical threshold on the capacity for a multimode channel.

\subsubsection{Shared states between multiple senders and a receiver}
\label{subsubsec:cv_multi_state}

\begin{figure}
    \centering
    \includegraphics[width=\linewidth]{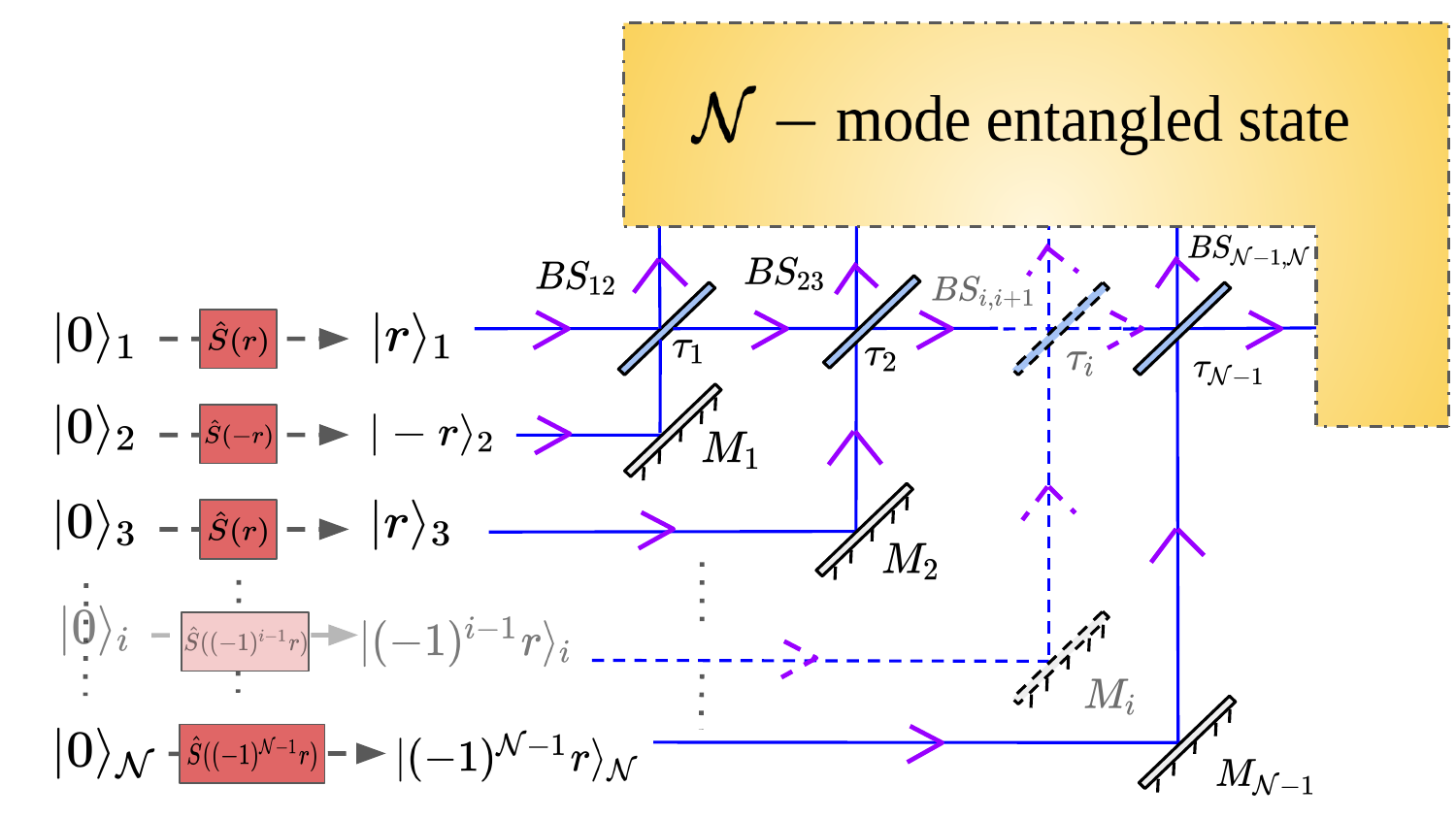}
    \caption{\textcolor{black}{Schematics to generate  an $\mathcal{N}$-mode entangled state to be used for dense coding between $\mathcal{N}-1$ senders and a single receiver. Initially, we start with an $\mathcal{N}$-mode vacuum state denoted as $|0\rangle_1 |0\rangle_2 \dots |0\rangle_{\mathcal{N}}$. Each mode is squeezed by a degree $r$, with alternate modes being squeezed in different quadratures. Specifically, modes undergoing momentum squeezing are denoted as $|r\rangle$ and are prepared by acting $S(r)$ on the initial vacuum mode while modes squeezed in the position quadrature are denoted as $|(-1)^{i-1}r\rangle$ which are prepared by acting $S((-1)^{i-1}r)$ on the vacuum mode. Further, the modes are combined pairwise using beam splitters i.e., modes $i$ and $i+1$ are combined using $BS_{i,i+1}$ of transmittivity $\tau_i$. This results in the $\mathcal{N}$-mode entangled state. In the figure, $M_i$s represent mirrors.}} 
    \label{fig:schematics2}
\end{figure}

We consider an $(\mathcal{N}-1)$-parameter family of $\mathcal{N}$-mode entangled states for the DC network between $(\mathcal{N}-1)$ senders, denoted as \(\mathcal{S}_1, \mathcal{S}_2, \ldots \mathcal{S}_{\mathcal{N}-1}\), and a single receiver, \(\mathcal{R}\). We now briefly mention a preparation procedure for such states starting from  single-mode squeezed states and linear optical elements, namely, the beam splitters.
In particular, we start from  $\mathcal{N}$ single-mode squeezed states of identical squeezing strengths $r$ but with alternately squeezed quadratures. These modes are entangled by a pairwise action of $(\mathcal{N} - 1)$ beam splitters with transmission coefficients, $\tau_1, \tau_2,...,$ and \(\tau_{\mathcal{N}-1}\) which leads to a  family of $(\mathcal{N}-1)$-parameter  genuinely multimode entangled states which serve as resources for distributed dense coding between  $(\mathcal{N}-1)$ senders and a single receiver. The generation of the resource states is schematically depicted in Fig. \ref{fig:schematics2}.
 The entanglement between the senders and the receiver (in the $(\mathcal{N}-1):1$ bipartition) depends upon the values of the parameters $\tau_i$ and we will show that the dense coding capacity does so too. 

\subsubsection{Encoding and Decoding}
\label{subsubsec:cv_multi_encoding_decoding}

The aim of the DC scheme is to transmit classical messages via an $\mathcal{N}$-mode entangled state which is  distributed between  $(\mathcal{N}-1)$ senders and the lone receiver. In particular, the protocol allows us to transmit $\mathcal{N}$ real numbers (which constitute the classical message) through this state. Suppose the sender, $i$,  encodes the classical message $\alpha_i$ in his/her mode with the help of a suitable displacement operator, $\hat{D}(\alpha_i)$. Note that the $\alpha_i$s are, in general, complex. Since we attempt to send only $\mathcal{N}$ real numbers, all but one choose the $\{\alpha_i \}$ to be real. Without loss of generality, we assume that the first sender encodes messages in both his/her input quadratures, i.e., $\alpha_1$ is chosen to be complex while
the remaining senders encode a single message, i.e., a real number which is in either the position or the momentum quadrature of the available mode. Thus, we have $\mathcal{N}$ encoded messages. Each sender encodes $\alpha_i$ from a Gaussian distribution of zero mean and standard deviation $\sigma$. Since LOCC is allowed between the senders, we can assume that the standard deviation is fixed among all the senders. 
The probability distribution of the input messages reads as
\begin{equation}
    p(\alpha) = \frac{1}{(2\pi \sigma^2)^{\mathcal{N}/2}} \exp \Big[-\sum_{i = 1}^\mathcal{N}\frac{\alpha_i^2}{2\sigma^2}\Big].
    \label{eq:cv_multi_p-alpha}
\end{equation}
Upon encoding, the senders' modes are transmitted to the receiver along a noiseless quantum channel. The receiver then applies  $(\mathcal{N}-1)$ beam splitters to combine the modes in a pairwise manner to start the decoding process. Therefore, the decoding essentially comprises the action of $\Big[\hat{B}_{(\mathcal{N}-1)\mathcal{N}}(\tau_{\mathcal{N}-1})...\hat{B}_{23}(\tau_2) \hat{B}_{12}(\tau_1)\Big]^\dagger$ on all the modes with \(\hat{B}_{ij}\) being the action of the beam splitters combining the modes, \(i\) and \(j\). This is followed by the homodyne measurements of suitable quadratures which are  performed to estimate the messages encoded by the senders. 
The decoding process  yields the conditional probability distribution $p(\beta|\alpha)$ where $\beta$ stands for the messages interpreted by the receiver upon decoding. The unconditional probability distribution of the decoded messages is then computed as 
\begin{equation}
p(\beta) = \int d^\mathcal{N} \alpha~ p(\beta|\alpha) p(\alpha),
\label{eq:p_beta_gen}
\end{equation}
and the mutual information quantifying the information achievable from the $\mathcal{N}$-mode states at the receiver's  side is given by
\begin{equation}
    \mathcal{I}(\mathcal{S}_1 \dots \mathcal{S}_{\mathcal{N}-1}:\mathcal{R}) = \int d^\mathcal{N} \alpha~d^\mathcal{N} \beta ~ p(\beta|\alpha) p(\alpha) \ln \Big[\frac{p(\beta|\alpha)}{p(\beta)}\Big].
    \label{eq:cv_multi_mutinfo}
\end{equation}
\textcolor{black}{Maximizing Eq. \eqref{eq:cv_multi_mutinfo} with respect to $\sigma$ under the constraint that the total number of photons  at  the modes of $(\mathcal{N}-1)$ senders is fixed to $\bar{N}$}, we obtain the capacity. We observe that for an $\mathcal{N}$-mode state, the total photon number of the senders' modes after encoding is given by
\begin{equation}
    \bar{N} = (\mathcal{N}-1) \sinh^2 r + \mathcal{N} \sigma^2,
    \label{eq:cv_multi_nbar}
\end{equation}
and the capacity of dense coding reads
\textcolor{black}{
\begin{eqnarray}
\nonumber C^{\mathcal{S}_1 \dots \mathcal{S}_{\mathcal{N} - 1}: \mathcal{R}}(\tau_1 \dots \tau_{\mathcal{N} - 1}) = && \max_{\substack{\sigma \\ \sum_{i = 1}^{\mathcal{N}-1} \bar{n}_i = \bar{{N}}}} \mathcal{I} (\mathcal{S}_1 \dots \mathcal{S}_{\mathcal{N}-1}:\mathcal{R}),\\
\end{eqnarray}
where the constraint involved in the maximization routine is $\sum_{i = 1}^{\mathcal{N}-1} \bar{n}_i = \bar{N}$.
}\\
The mutual information is optimized when
\begin{equation}
    \bar{N} = (\mathcal{N} - 1) e^r \sinh r.
    \label{eq:cv_multi_opt_condn}
\end{equation}
Choosing $\sigma$ as $(\frac{\mathcal{N}-1}{2\mathcal{N}} \sinh{2r})^{1/2}$ by substituting  Eq. \eqref{eq:cv_multi_opt_condn} in Eq. \eqref{eq:cv_multi_nbar} for a given $\mathcal{N}$-mode state with sender signal strength $\bar{N}$, we can find the classical  capacity of the quantum channel.


\subsubsection{Classical threshold}

The advantage of a quantum protocol in dense coding is assured if its capacity surpasses that of the corresponding classically available scheme. Therefore, we need to set a benchmark  with which the classical capacity of a quantum channel can be compared. According to Holevo's theorem, if a classical message, say $\alpha$, taken from a probability distribution $p(\alpha)$ is to be transmitted via a quantum state $\hat{\rho}_\alpha$,  the mutual information $\mathcal{I}(\mathcal{S}:\mathcal{R})$ between the sender, \(\mathcal{S}\) and the receiver, \(\mathcal{R}\)  is bounded above by the Holevo quantity \cite{Holevo-1998}, 
\begin{equation}
    \mathcal{I} (\mathcal{S}: \mathcal{R}) \leq S(\hat{\rho}) - \int~ d^2\alpha~ p(\alpha)~ S(\hat{\rho}_\alpha) \leq  S(\hat{\rho}),
    \label{eq:Holevo_bound_CV}
\end{equation}
where $S(\hat{\rho}) = -\text{tr}(\hat \rho\ln \hat \rho)$, is the von Neumann entropy of the density operator $\hat{\rho}=\int~ d^2\alpha~ p(\alpha)~\hat{\rho}_\alpha$.

Considering a legitimate constraint of having a fixed mean number of photons $\bar{N}$ (which can be modulated), the required task is to find the configuration of a single-mode bosonic field in order to maximize the mutual information, $\mathcal{I}(\mathcal{S}:\mathcal{R})$. It was shown \cite{drummond_Caves-1994,Yuen_Ozawa-1993} that the optimal channel capacity via the classical protocol is achieved by photon counting measurement from an ensemble of number states having maximum entropy, i.e., $\sum_n P(n)\ket{n}\bra{n}$ with $P(n)=\bar{N}^n(1+\bar{N})^{-(n+1)}$.

With this optimal configuration of a single-mode bosonic channel, the channel capacity for a single sender and a single receiver without entanglement  is found to be \cite{Yuen_Ozawa-1993}
\begin{equation}
    C_{cl}^{\mathcal{S}:\mathcal{R}} (\bar{N})=(1+\bar{N})\ln(1+\bar{N})-\bar{N}\ln\bar{N} .
    \label{eq:singlemode_classical_capacity}
\end{equation}
In a similar spirit, the capacity  with $\mathcal{N}-1$ senders and a single  receiver yields
\begin{equation}
    C_{cl}^{\mathcal{S}_1 \ldots \mathcal{S}_{\mathcal{N}-1}:\mathcal{R}} (\{\bar{N}_i\})=\sum_i^{\mathcal{N}-1}[(1+\bar{N}_i)\ln(1+\bar{N}_i)-\bar{N}_i\ln\bar{N}_i] ,
    \label{eq:Nmode_classical_capacity}
\end{equation}
where $\bar{N}_i$ is the mean photon number of the  sender's mode, \(i\).
Imposing the constraint of having a fixed mean photon number $\bar{N}$ at the senders' mode, where $\bar{N}=\sum_i^{(\mathcal{N}-1)}\bar{N}_i$,  the capacity in the classical scenario where entanglement between senders and a receiver is absent can be obtained by maximizing $ C_{cl}^{\mathcal{S}_1 \ldots \mathcal{S}_{\mathcal{N}-1}:\mathcal{R}} (\{\bar{N}_i\})$ over $\{ \bar{N}_i\}$ with the constraint $\bar{N}=\sum_i^{(\mathcal{N}-1)}\bar{N}_i$.
The condition for achieving the maximum capacity turns out  to be $\bar{N_i}=\bar{N}/(\mathcal{N}-1)$ with equal distribution of photons being taken at all senders' modes. Substituting $\bar{N_i}=\bar{N}/(\mathcal{N}-1)$ in Eq. (\ref{eq:Nmode_classical_capacity}), we obtain the expression for  capacity in the classical case with an arbitrary number of senders and a single receiver as
\begin{eqnarray}
    C_{cl}^{\mathcal{S}_1 \ldots \mathcal{S}_{\mathcal{N}-1}:\mathcal{R}} =&& (\mathcal{N}-1)\Big[(1+\frac{\bar{N}}{\mathcal{N}-1})\ln(1+\frac{\bar{N}}{\mathcal{N}-1})-\nonumber\\
    &&\frac{\bar{N}}{\mathcal{N}-1}\ln\frac{\bar{N}}{\mathcal{N}-1}\Big].
    \label{eq:Nmode_classical_capacity_final}
\end{eqnarray}
Comparing  \(C_{cl}^{\mathcal{S}_1 \ldots \mathcal{S}_{\mathcal{N}-1}:\mathcal{R}}\) with the capacity obtained via a shared entangled state, we can confirm the quantum advantage which we will demonstrate explicitly for the shared three- and four-mode states in the succeeding sections.

\section{Classical Capacity for three-mode channel involving two senders and a single receiver}
\label{sec:3modecap}

To derive the expression for the classical capacity between two senders, and a  single receiver,  $\mathcal{S}_1,  \mathcal{S}_2$  and  $\mathcal{R}$ respectively,  a  three-mode squeezed state is initially distributed among them.  A three-mode  genuinely  multimode entangled state is, in general,  prepared with the help of a tritter. The class of such states constitutes a two-parameter family, characterized by the transmittivities, $\tau_1$ and $\tau_2$ of two beam splitters that comprise the tritter. The three-mode entangled state  identified by its displacement vector and covariance matrix can be represented as
\begin{eqnarray}
&&\textbf{d}_{0} = (0,0,0,0,0,0)^T \label{eq:d3msv}\\
&&\Xi_0 = \left(
\begin{array}{cccccc}
 \mathcal{A} & 0 & \mathcal{R} & 0 & \mathcal{T} & 0 \\
 0 & \mathcal{B} & 0 & -\mathcal{R} & 0 & -\mathcal{T} \\
 \mathcal{R} & 0 & \mathcal{C} & 0 & -\mathcal{S} & 0 \\
 0 & -\mathcal{R} & 0 & \mathcal{D} & 0 & \mathcal{S} \\
 \mathcal{T} & 0 & -\mathcal{S} & 0 & \mathcal{E} & 0 \\
 0 & -\mathcal{T} & 0 & \mathcal{S} & 0 & \mathcal{F} \\
\end{array}
\right)
\label{eq:sigma_3msv}
\end{eqnarray}
where 
\begin{eqnarray}
\nonumber &&\mathcal{A} = \frac{1}{2} e^{-2 r} [\left( e^{4 r}-1\right) \tau_1 +1], \label{eq:sigma0_A}\\
\nonumber &&\mathcal{B} = \frac{1}{2}\left( e^{-2 r} \tau_1 + e^{2 r} (1-\tau_1)\right), \label{eq:sigma0_B} \\
\nonumber &&\mathcal{C} = \frac{1}{2} \left(\sinh 2 r  (1 -2 \tau_1 \tau_2)+ \cosh 2 r\right), \label{eq:sigma0_C} \\
\nonumber &&\mathcal{D} = \frac{1}{2} e^{-2 r} [\left( e^{4 r}-1\right) \tau_1 \tau_2+1], \label{eq:sigma0_D} ~~\\
\nonumber &&\mathcal{E} = \frac{1}{2} \left( \sinh 2 r (1-2 \tau_1 (1-\tau_2))+ \cosh 2 r\right), \label{eq:sigma0_E}\\
\nonumber && \mathcal{F} =\frac{1}{2} e^{-2 r} [1+\tau_1 (e^{4 r} -1 )(1 - \tau_2)],  \\
\nonumber && \mathcal{R} = \sqrt{ \tau_1 \tau_2 (1-\tau_1)} ~\sinh 2 r , \label{eq:sigma0_R} \\
\nonumber && \mathcal{S} = \tau_1 \sqrt{\tau_2 (1-\tau_2) } ~\sinh 2 r , \label{eq:sigma0_S} \\
&& \mathcal{T} = \sqrt{\tau_1 (1-\tau_1) (1-\tau_2) } ~\sinh 2 r.
\label{eq:sigma0_T}
\end{eqnarray}
All the initial single mode squeezed states  are considered to have equal squeezing strength, $r$. For $\tau_1 = 1/3$ and $\tau_2 = 1/2$, we obtain the well-known basset-hound state \cite{Braunstein_PRL_2000,Adesso_arXiv_2006,Adesso_JPA_2007}. 

{\it Encoding by the senders.} 
Since the state comprises three modes, two senders can send at most three real numbers accurately. Without loss of generality, we assume that $\mathcal{S}_1$ sends two real numbers $\alpha_{1x}$ and $\alpha_{1y}$, encoded through a suitable displacement operation, $\hat{\mathcal{D}}_1(\alpha)$ where $\alpha_1 = \alpha_{1x} + i \alpha_{1y}$ while $\mathcal{S}_2$ chooses to send a single real number $\alpha_{2y}$ with the help of the displacement,  $\hat{\mathcal{D}}_2(\alpha_2)$ having $\alpha_{2} = i \alpha_{2y}$. Both the senders resort to a Gaussian distribution of their respective real numbers, having the 
same standard deviation $\sigma$. The input probability distribution is then given by
\begin{equation}
    p(\alpha) = \frac{1}{2\pi \sigma^2} \exp(- \frac{|\alpha_1|^2}{2\sigma^2}) \frac{1}{\sqrt{2\pi} \sigma} \exp(- \frac{|\alpha_2|^2}{2\sigma^2}).
    \label{eq:input_prob}
\end{equation}
The encoding process gives rise to the  displacement vector and covariance matrices, given by
\begin{eqnarray}
&&\textbf{d}_{en} = (\sqrt{2} \alpha_{1x}, \sqrt{2} \alpha_{1y}, 0 , \sqrt{2} \alpha_{2y}, 0, 0)^T, \label{eq:encoded_disp}\\
&&~~~~~~~~~~~~~~~~~~~\Xi_{en} = \Xi_0.
\label{eq:encoded_sigma}
\end{eqnarray}

{\it Decoding by the receiver.} After the encoding process, senders send their respective modes to the receiver, and hence the  receiver possesses  the three-mode state. Towards recovering the classical information,  
two beam splitters are used to combine modes \(1\) and \(2\) as well as modes \(2\) and \(3\),  $\Big[\hat{B}_{23}(\tau_2) \hat{B}_{12}(\tau_1)\Big]^\dagger$. Such a decoding routine results in a three-mode state with the  displacement vector and covariance matrix, respectively as
\begin{eqnarray}
&&\textbf{d}_{dec}   = (\sqrt{2\tau_1}\alpha_{1x}, \sqrt{2 \tau_2 (1-\tau_1)} \alpha_{2y} + \sqrt{2\tau_1}\alpha_{1y},\nonumber\\ &&\sqrt{2(1-\tau_1)}\alpha_{1x}, \sqrt{2(1-\tau_1)}\alpha_{1y} - \sqrt{2\tau_1\tau_2}\alpha_{2y},\nonumber \\ 
&& 0, \sqrt{2(1-\tau_2)}\alpha_{2y})^T, ~~~~~~~~~~~~~ 
\label{eq:decoded_disp} \\
&&\Xi_{dec}  = \text{diag}(\frac{1}{2}e^{2r},\frac{1}{2}e^{-2r},\frac{1}{2}e^{-2r},\frac{1}{2}e^{2r},\frac{1}{2}e^{2r},\frac{1}{2}e^{-2r}).\nonumber \\ \label{eq:decoded_sigma}
\end{eqnarray}
The receiver requires to undertake a homodyne detection to measure $p_1, x_2$, and $p_3$, since these quantities have the lowest variance in $\Xi_{dec}$. 
It results in the probability distribution of the output variables (conditioned on the input) as
\begin{eqnarray}
    \nonumber && p(\beta|\alpha)  = \int dx_1 dp_2 dx_3 W_{\rho_{\mathcal{S}_1 \mathcal{S}_2 \mathcal{R}}}(x_1,\beta_1,\beta_2,p_2:x_3,\beta_3)\\
    &&  = \frac{1}{\pi^{3/2}} \Big[ \exp \{ 3r - e^{2r}(\beta_2 - \alpha_{1x}\sqrt{2(1-\tau_1)})^2\}\Big] \times \nonumber \\
    && \Big[ \exp \{ - e^{2r}(\beta_1  - \sqrt{2} (\alpha_{1y}\sqrt{\tau_1} + \alpha_{2y}\sqrt{\tau_2(1-\tau_1)}))^2\}\Big] \times \nonumber \\
    && \Big[ \exp \{ -e^{2r} (\beta_3 - \alpha_{2y}\sqrt{2(1-\tau_2)})^2\} \Big]
    \label{eq:p_beta_alpha}
\end{eqnarray}
where $W_{\rho_{\mathcal{S}_1 \mathcal{S}_2 \mathcal{R}}}$ is the Wigner function of the state after the modes are combined by the receiver using the beam splitter setup described above \cite{Lee_PRA_2014}. $\beta_i \,\, (i=1, 2, 3)$ represent the homodyne outcomes obtained by the receiver upon measuring on the mode, $i$.
The unconditioned probability of the homodyne variables from Eq. \eqref{eq:p_beta_gen} in this case reads
\begin{equation}
    p(\beta) = \int d^2 \alpha_1 d \alpha_2 p (\beta| \alpha) p(\alpha).
    \label{eq:p_beta}
\end{equation}
Using Eqs. \eqref{eq:input_prob} - \eqref{eq:p_beta}, the mutual information corresponding to this channel can be computed as 
\begin{eqnarray}
 \mathcal{I}(\mathcal{S}_1 \mathcal{S}_2&&: \mathcal{R}) = \int d^3 \beta d^2 \alpha_1 d \alpha_2 p(\beta|\alpha) p(\alpha) \ln \Big[\frac{p(\beta|\alpha)}{p(\beta)}\Big] ~~~~~~~~\nonumber \\
 &&= \frac{1}{2} \ln \Big[ \left(4 e^{2 r} \sigma^2+1\right) \left(4 e^{2 r} \sigma^2 (1-\tau_1)+1\right)\Big] +  \nonumber \\
 && ~~~~\frac{1}{2} \ln \Big[ \left(4 e^{2 r} \sigma^2 \tau_1 (1-\tau_2)+1\right)\Big]. 
 \label{eq:mut_info}
\end{eqnarray}
\textcolor{black}{Since the decoding scheme is fixed to homodyne detection, the dense coding capacity is obtained by maximizing Eq. \eqref{eq:mut_info} over the standard deviation of the encoding displacement operations subject to a fixed average photon number constraint.} This condition can be represented as
\begin{equation}
    \Bar{n}_1 + \Bar{n}_2 = 2 \sinh^2{r} + 3 \sigma^2 = \bar{N}.
\end{equation}
For a fixed $\Bar{N}$, the mutual information is maximized when $\sigma^2=\frac{1}{3}\sinh{2r}$ and $r = (1/2)\ln(1 + \Bar{N})$, leading to the expression for the dense coding capacity,
\textcolor{black}{
\begin{widetext}
\begin{eqnarray}
\nonumber C^{\mathcal{S}_1 \mathcal{S}_2: \mathcal{R}}(\tau_1, \tau_2) = && \max_{\substack{\sigma \\ \bar{n}_1 + \bar{n}_2 = \bar{\mathcal{N}}}} \mathcal{I} (\mathcal{S}_1 \mathcal{S}_2 : \mathcal{R})\\
 = && \frac{1}{2} \ln \Big[\frac{1}{27} (2 \bar{N} (\bar{N}+2)+3) (2 \bar{N} (\bar{N}+2) (1 - \tau_1)+3) (2 \bar{N} (\bar{N}+2) \tau_1 (1 - \tau_2)+3)\Big].
\label{eq:capacity}
\end{eqnarray}
\end{widetext}
}
Substituting various values of $\tau_1$ and $\tau_2$, we obtain the CVDC capacity for different states belonging to the two-parameter family. Notice that  although the basset-hound state obtained with $\tau_1 = 1/3$ and $\tau_2 = 1/2$, possesses the maximum genuine multimode entanglement in this set of states,  
we find that there exist states (obtained with other values of $\tau_1$ and $\tau_2$) which furnish a greater CV dense coding capacity than that obtained via basset-hound state (cf. \cite{Tamoghna}).
For example, with $\tau_1 = \tau_2 = 1/2$,
the DC capacity takes the form as
\begin{widetext}
\begin{equation}
    C^{\mathcal{S}_1 \mathcal{S}_2: \mathcal{R}}(1/2,1/2) = \frac{1}{2} \ln \Big[\frac{1}{54} (\bar{N} (\bar{N}+2)+3) (\bar{N} (\bar{N}+2)+6) (2 \bar{N} (\bar{N}+2)+3)\Big],
    \label{eq:3mode_cap_tau_half}
\end{equation}
\end{widetext}
which increases monotonically  with the increase of \(\bar{N}\). and for a given \(\bar{N}\), we notice that  \(C^{\mathcal{S}_1 \mathcal{S}_2: \mathcal{R}}(1/2,1/2) > C^{\mathcal{S}_1 \mathcal{S}_2: \mathcal{R}}(1/3,1/2)\). 

\subsection{Quantum advantage in DC}
\label{subsec:3mode_classical}

\begin{figure}
    \centering
    \includegraphics[width = \linewidth]{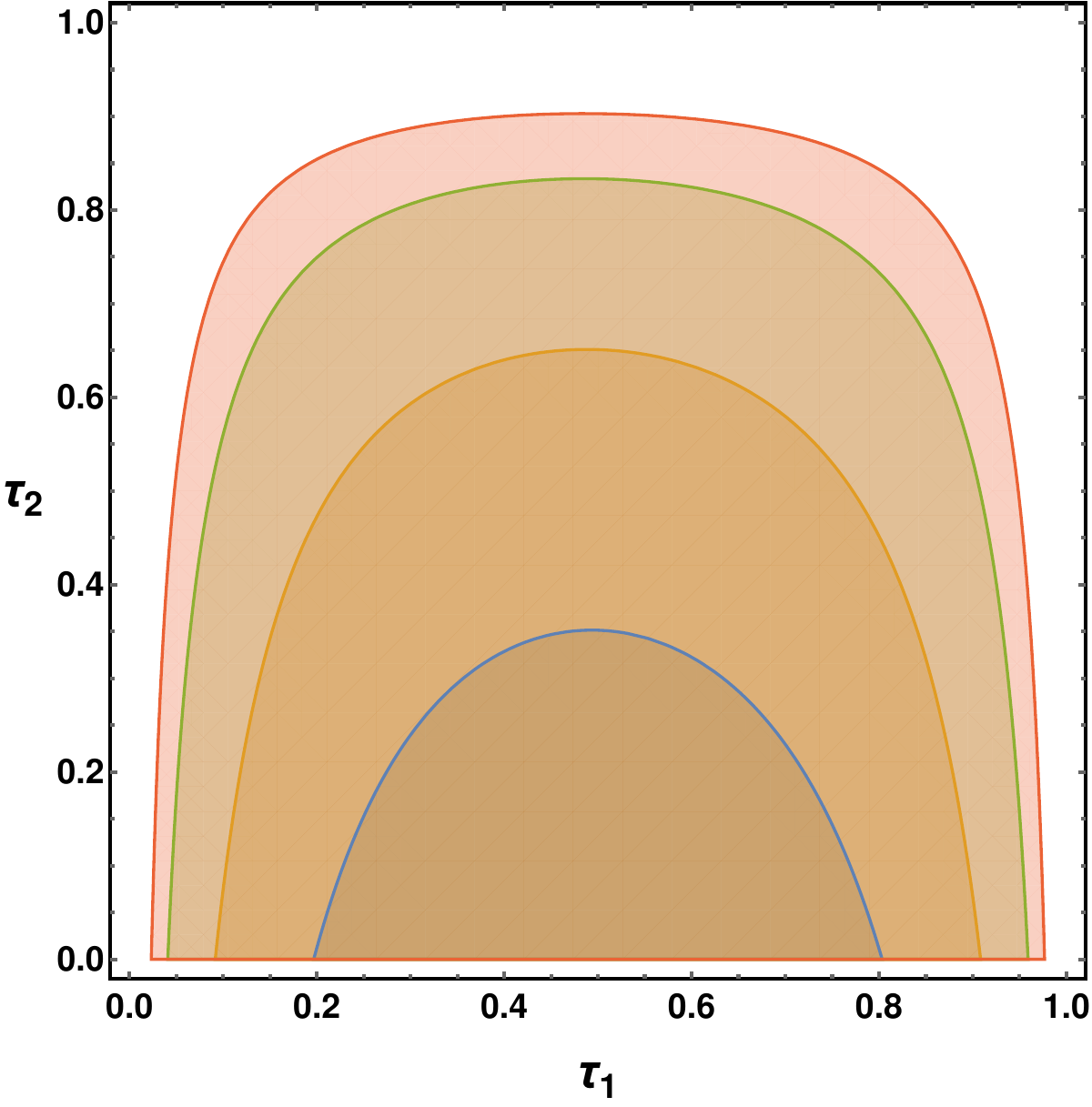}
    \caption{Region plot of the state space  bounded by $\tau_1$ and $\tau_2$ which provide the quantum advantage of DC (in bits) having two senders and a single receiver with states belonging to the two-parameter family of the three-mode states defined in the main text. The abscissa and ordinate represent $\tau_1$ and  $\tau_2$ respectively. 
    The blue, orange, green, and red curves (from below) correspond to $\bar{N} = 7$,  $\bar{N} = 10$, $\bar{N} = 15$ and $\bar{N} = 20$. Note that all the states bounded by each curve can provide a quantum advantage in DC.   All the axes are dimensionless. }
    \label{fig:3mode_region}
\end{figure}

To guarantee the quantum advantage, it is important to compare the classical capacity of a quantum channel with the capacity in a classical protocol. From  Eq. \eqref{eq:Nmode_classical_capacity_final}, 
the optimum  capacity in the classical case for a channel with mean photon number $\bar{N}$ shared between two senders and one receiver reduces to 
\begin{equation}
    C_{cl}^{\mathcal{S}_1 \mathcal{S}_2: \mathcal{R}} = 2(1 + \bar{N}/2) \ln (1 + \bar{N}/2) - 2 (\bar{N}/2) \ln (\bar{N}/2).
    \label{eq:3mode_c-cap}
\end{equation}
Let us define the quantum advantage in the DC network involving an arbitrary number of senders and a single receiver  as
\begin{eqnarray}
&&\Delta^{\mathcal{S}_1 \mathcal{S}_2 \cdots \mathcal{S}_{\mathcal{N}-1}: \mathcal{R}} =  \nonumber \\ 
 && C^{\mathcal{S}_1 \mathcal{S}_2 \mathcal{S}_{\mathcal{N}-1}: \mathcal{R}}(\tau_1, \tau_2, \cdots, \tau_{\mathcal{N}-1}) - C_{cl}^{\mathcal{S}_1 \mathcal{S}_2 \cdots \mathcal{S}_{\mathcal{N}-1}: \mathcal{R}}, 
 \label{eq:qa}
\end{eqnarray}
for a fixed photon number. The positivity of the above ensures quantum advantage in the shared channels. 

Let us identify the range of $\tau_1$ and $\tau_2$ for which the three-mode state provides a quantum advantage, i.e., \(\Delta^{\mathcal{S}_1 \mathcal{S}_2: \mathcal{R}} >0\) for a fixed $\bar{N}$ as illustrated in Fig. \ref{fig:3mode_region}. We find that, with increasing $\bar{N}$, the region bounded in the  $(\tau_1,\tau_2)$-plane providing quantum advantage also grows in size. Furthermore,  with $\tau_1 = 0.5$, we find the largest range of $\tau_2$ which provides a quantum advantage for a given $\bar{N}$. This indicates that states prepared with $\tau_1 = 0.5$ are more suitable for multimode DC between two senders and a lone receiver.

{\it Threshold energy for quantum advantage.} For any given values of $\tau_1$ and $\tau_2$, there exists a threshold energy, say, $\bar{N}^{\mathcal{S}_1 \mathcal{S}_2: \mathcal{R}}_{th}(\tau_1,\tau_2)$, above which the quantum advantage can be achieved. Although it is very hard to find such analytical expression of $\bar{N}^{\mathcal{S}_1 \mathcal{S}_2: \mathcal{R}}_{th}(\tau_1,\tau_2)$, we can find the threshold energy numerically for a given $\tau_1$ and $\tau_2$ by solving the equation $C^{\mathcal{S}_1 \mathcal{S}_2:\mathcal{R}}(\tau_1,\tau_2)=C^{\mathcal{S}_1 \mathcal{S}_2: \mathcal{R}}_{cl}$ for $\bar{N}$. For example, we find the value of $\bar{N}^{\mathcal{S}_1 \mathcal{S}_2: \mathcal{R}}_{th}(1/2,1/2) = 8.15$, i.e., the three-mode entangled state having the state parameters $\tau_1 = \tau_2 = 1/2$ can offer a quantum advantage in the DC protocol at the minimum expense of energy $\bar{N}^{\mathcal{S}_1 \mathcal{S}_2: \mathcal{R}}_{th}(1/2,1/2) = 8.15$. 
For a given energy $\bar{N}$, we find that
\begin{eqnarray}
\tau_1\big|_{\substack{\max\\\min}} &&=  0.5 \pm \nonumber\\ 
&&\frac{\sqrt{16 (\bar{N} (\bar{N}+2)+3)^2-\frac{27 \bar{N}^{-2 \bar{N}} (\bar{N}+2)^{2 \bar{N}+4}}{2 \bar{N} (\bar{N}+2)+3}}}{8 \bar{N} (\bar{N}+2)},
\end{eqnarray}

and for a given $\tau_1$ and $\bar{N}$,
\begin{eqnarray}
\tau_2\big|_{{\max}} && = 1 + \frac{3}{2(\bar{N}^2 {\tau_1}+2 \bar{N} {\tau_1})}- \nonumber \\
&&\frac{27 \bar{N} ^{-2 \bar{N}-1} (\bar{N}+2)^{2 \bar{N}+3}}{32 {\tau_1} (2 \bar{N} (\bar{N}+2)+3) (2 \bar{N} (\bar{N}+2) (1-{\tau_1})+3)}. \nonumber\\
\end{eqnarray}
where $\tau_i\big|_{\substack{\min\\\max}}$ represent the region bounded by $\tau_i$ which provides quantum advantage.
Moreover, the minimum number of photons at senders' mode required to avail the quantum advantage is then given by
\begin{equation*}
    \bar{N}^{\mathcal{S}_1 \mathcal{S}_2: \mathcal{R}}_{th}\big|_{\min\limits_{\{\tau_1,\tau_2\}}} = 5.38,
\end{equation*}
where minimization is performed over all possible values of \(\tau_1\) and \(\tau_2\).

{\it Quantum advantage with large squeezing strength.} Let us now investigate the ratio of the classical capacity of a quantum channel and  the  capacity in the classical protocol for large resource squeezing $r$. 
Substituting $\bar{N}=2e^{r}\sinh{r}$ (see Eq. (\ref{eq:cv_multi_opt_condn})) into Eq. (\ref{eq:capacity}), we obtain $C^{\mathcal{S}_1 \mathcal{S}_2: \mathcal{R}}\sim 6r$, whereas the same substitution in Eq. (\ref{eq:3mode_c-cap}) yields $C^{\mathcal{S}_1 \mathcal{S}_2: \mathcal{R}}_{cl}\sim4r$ for large $r$. Hence the ratio becomes
\begin{equation}
   \frac{C^{\mathcal{S}_1 \mathcal{S}_2: \mathcal{R}}}{C^{\mathcal{S}_1 \mathcal{S}_2: \mathcal{R}}_{cl}} = \frac{3}{2} \;\;\;\;\;\;\text{(at large $r$)}.
   \label{eq:ratio_3}
\end{equation}

 Knowing that the quantum protocol for $\tau_1 = \tau_2 = 1/2$ can overcome the classical threshold value,  when the total photon number of the senders' modes is $\bar{N} \geq 8.15$, we can find that the minimum squeezing required for quantum advantage in the two sender-one receiver scenario, denoted by $r_{break-even}^{\mathcal{S}_1 \mathcal{S}_2: \mathcal{R}} (\tau_1=1/2, \tau_2=1/2)$ is  $1.10685$. 
 Note, however, that $r_{break-even}^{\mathcal{S}_1\mathcal{S}_2:\mathcal{R}} (1/2, 1/2)$ is higher than that for the single sender-single receiver regime \cite{Braunstein_PRA_2000}. It is due to the fact that $C_{cl}^{\mathcal{S}_1 \mathcal{S}_2: \mathcal{R}}$ is much higher than the classical bound for the DC protocol with a single sender-receiver duo. 

\section{Multimode Dense coding network  with four-mode states}
\label{sec:four-mode-dc}

Akin to the case for three-mode channels, let us consider a general class of four-mode genuinely entangled Gaussian states, characterized by three parameters, $\tau_1, \tau_2$ and $\tau_3$, shared between three senders, \(\mathcal{S}_i\), (\(i=1, 2, 3\)) and a receiver, \(\mathcal{R}\). 

{\it Encoding.} 
Three senders, $\mathcal{S}_1, \mathcal{S}_2$, and $\mathcal{S}_3$, perform displacement operators on their respective modes as a part of the encoding process. The displacement amplitude for each sender is proportional to the message they wish to send. Like in the previous three-mode situation, we assume, without loss of generality, that $\mathcal{S}_1$ incorporates displacement in both the quadratures of his/her available mode with an amplitude $\alpha_1 = \alpha_{1x} + i \alpha_{1y}$. $\mathcal{S}_2$ chooses to displace only the momentum quadrature by $\alpha_{2y}$ while the position displacement $\alpha_{3x}$ is performed by  $\mathcal{S}_3$. The input messages belong to a Gaussian ensemble characterized by the  probability distribution,
\begin{equation}
    p(\alpha) = \frac{1}{(2\pi \sigma^2)^2} \exp \Big[ - \frac{1}{2\sigma^2} (\alpha_{1x}^2 + \alpha_{1y}^2 + \alpha_{2y}^2 + \alpha_{3x}^2) \Big].
    \label{eq:4mode-palpha}
\end{equation}
The senders then transfer their modes, post-encoding, to the receiver $\mathcal{R}$ via noiseless quantum channels.

{\it Decoding.} In order to decode the messages, the receiver combines all the four modes at his disposal, with the help of the  beam splitter setup, represented as $$\Big[\hat{B}_{34}(\tau_3)\hat{B}_{23}(\tau_2)\hat{B}_{12}(\tau_1)\Big]^\dagger.$$ The homodyne detection by the receiver on modes $p_1, x_2, p_3$ and $x_4$ leads to the conditional probability on the decoded message (here, the subscript on the numbers indicates the quadrature on which the homodyne detection is performed), given by 
\begin{widetext}
\begin{eqnarray}
    \nonumber  p(\beta|\alpha) &&= \int dx_1 dp_2 dx_3 dp_4 W_{\rho_{\mathcal{S}_1\mathcal{S}_2\mathcal{S}_3\mathcal{R}}}(x_1,\beta_1,\beta_2,p_2,x_3,\beta_3:\beta_4,p_4) \\
    \nonumber  = && \frac{1}{\pi} \exp (4 r-e^{2 r} \{[\beta_2-\sqrt{2} (\alpha_{1x} \sqrt{1-\tau_1}-\alpha_{3x} \sqrt{\tau_1 \tau_3 (1-\tau_2) }]^2+[\beta_1-\sqrt{2} (\alpha_{1y} \sqrt{\tau_1}+\alpha_{2y} \sqrt{(1-\tau_1) \tau_2})]^2\})  \\ \times
   && \frac{1}{\pi} \exp \{-e^{2r}[\left(\beta_3-\sqrt{2} \alpha_{2y}\sqrt{1-\tau_2}\right)^2 +\left(\beta_4-\sqrt{2} \alpha_{3x} \sqrt{1-\tau_3}\right)^2 ]\}.
\end{eqnarray}
\end{widetext}

Here, $ W_{\rho_{\mathcal{S}_1\mathcal{S}_2\mathcal{S}_3\mathcal{R}}}$ again represents the Wigner function of the state after the beam splitter operation  by the receiver and $\beta_i$ are the homodyne outcomes for the mode,  $i$.
Following the same steps as in the case of the three-mode states, one can calculate the unconditioned decoding probability distribution $p(\beta)$ using Eq. \eqref{eq:p_beta_gen}, whereafter, the mutual information  can be estimated as
\begin{widetext}
\begin{eqnarray}
&&\nonumber \mathcal{I}(\mathcal{S}_1\mathcal{S}_2\mathcal{S}_3:\mathcal{R}) = \int d^4 \beta d^2 \alpha_1 d \alpha_2 d \alpha_3 p(\beta|\alpha) p(\alpha) \ln \Big[\frac{p(\beta|\alpha)}{p(\beta)}\Big]  \\
 && \nonumber = \frac{1}{2} \ln \Big[\sigma^8 \left(e^{2 r} 4(1- \tau_1 \tau_2)+\frac{1}{\sigma^2}\right) \left(\frac{\left(4 e^{2 r} \sigma^2+1\right) \left(4 e^{2 r} \sigma^2 \tau_1 (1-\tau_2)+1\right)}{\sigma^2 \left(4 e^{2 r} \sigma^2 (1-\tau_1\tau_2)+1\right)}\right) \left(\frac{1}{\sigma^2}+4 e^{2 r} (\tau_1 (1-\tau_2) \tau_3+(1 - \tau_3))\right)\Big] + \\
 && \frac{1}{2}\ln \Big[\frac{16 e^{4 r} \sigma^4 (1-\tau_1) (1-\tau_3)+4 e^{2 r} \sigma^2 (\tau_1 (1-\tau_2) \tau_3-\tau_1-\tau_3+2)+1}{\sigma^2+4 e^{2 r} \sigma^4 (\tau_1 (1-\tau_2) \tau_3-\tau_3+1)}\Big].
 \label{eq:4mode_mut_info}
\end{eqnarray}
\end{widetext}
Optimization of Eq. \eqref{eq:4mode_mut_info} subject to a fixed photon number $\bar{N}$ at the senders' ends, i.e., $\bar{N} = \bar{n}_1 + \bar{n}_2 + \bar{n}_3 = 3 \sinh^2 r + 4 \sigma^2$ leads to the DC capacity of a network involving three senders and one receiver. With the aid of optimal conditions given by $\sigma^2=\frac{3}{8} \sinh{2r}$ and $r=(1/2)\ln{(1+\frac{2\bar{N}}{3})}$, we obtain the capacity in terms of the photon strength of the senders and the state parameters as 

\begin{widetext}
\begin{eqnarray}
  \nonumber C^{\mathcal{S}_1 \mathcal{S}_2 \mathcal{S}_3: \mathcal{R}}(\tau_1, \tau_2, \tau_3) &&= \max_{\substack{\sigma \\ \bar{n}_1 + \bar{n}_2 = \bar{\mathcal{N}}}} \mathcal{I} (\mathcal{S}_1 \mathcal{S}_2 \mathcal{S}_3: \mathcal{R})\\ = && \frac{1}{2}\ln \left(\frac{1}{81} [\bar{N} (\bar{N}+3)+3] [\bar{N} (\bar{N}+3) \tau _1 \left(1-\tau _2\right)+3]\right) + \nonumber\\&& \nonumber \frac{1}{2}\ln [\left([\bar{N} (\bar{N}+3)+3] [3+\bar{N} (\bar{N}+3) \left(1-\tau _1\right)]+\bar{N} (\bar{N}+3) \tau _3 \{[\bar{N} (\bar{N}+3)+3] \left(\tau _1-1\right)-3 \tau _1 \tau _2\}\right)]. \\
~~~~\label{eq:4mode_cap}
\end{eqnarray}
\end{widetext}

\begin{widetext}
{\it Classification of multimode states according to their DC capacities.} Motivated from the three-mode results, let us first consider a symmetric situation, i.e., when  $\tau_1 = \tau_2 = \tau_3 = 1/2$, the DC capacity becomes
\begin{eqnarray}
C^{\mathcal{S}_1\mathcal{S}_2\mathcal{S}_3:\mathcal{R}}(1/2,1/2,1/2) = \frac{1}{2} \ln \Big[(\bar{N} (\bar{N}+3)+3) (\bar{N} (\bar{N}+3)+12) (\bar{N} (\bar{N}+3) (2 \bar{N} (\bar{N}+3)+27)+72)\Big] - \ln[36 \sqrt{2}]. ~~ ~~~~~~  \label{eq:4mode_opt_cap}
\end{eqnarray}
\end{widetext}
\begin{widetext}
Instead of equal \(\tau_i\)s, let us choose \(\tau_1= 1/3 \), \(\tau_2=1/4 \), \(\tau_3=4/5 \), in which case the DC capacity reads as
\begin{equation}
C^{\mathcal{S}_1\mathcal{S}_2\mathcal{S}_3:\mathcal{R}}(1/3,1/4,4/5) = \frac{1}{2} \ln \Big[(\bar{N} (\bar{N}+3)+3) (\bar{N} (\bar{N}+3)+12) (2 \bar{N} (\bar{N}+3) (\bar{N} (\bar{N}+3)+24)+135)\Big] - \ln[18 \sqrt{15}] ~~~~~~~~~
\label{eq:4mode_cap_diff_tau}
\end{equation}
\end{widetext}
Comparing Eqs. \eqref{eq:4mode_opt_cap} and \eqref{eq:4mode_cap_diff_tau}, we find $$C^{\mathcal{S}_1\mathcal{S}_2\mathcal{S}_3:\mathcal{R}}(1/2,1/2,1/2) > 
C^{\mathcal{S}_1\mathcal{S}_2\mathcal{S}_3:\mathcal{R}}(1/3,1/4,4/5).$$ 
To demonstrate it more explicitly, we
 vary \(\tau_i\)s and find the hierarchies among states which are beneficial for classical information transmission by using Eq. \eqref{eq:4mode_cap} with a fixed photon number \(\bar{N}\) (see Fig. \ref{fig:4mode_region}). 

\subsection{Outperforming quantum network with  four-mode classical scheme}

\label{subsec:3mode_classical}

\begin{figure}
    \centering
    \includegraphics[width = \linewidth]{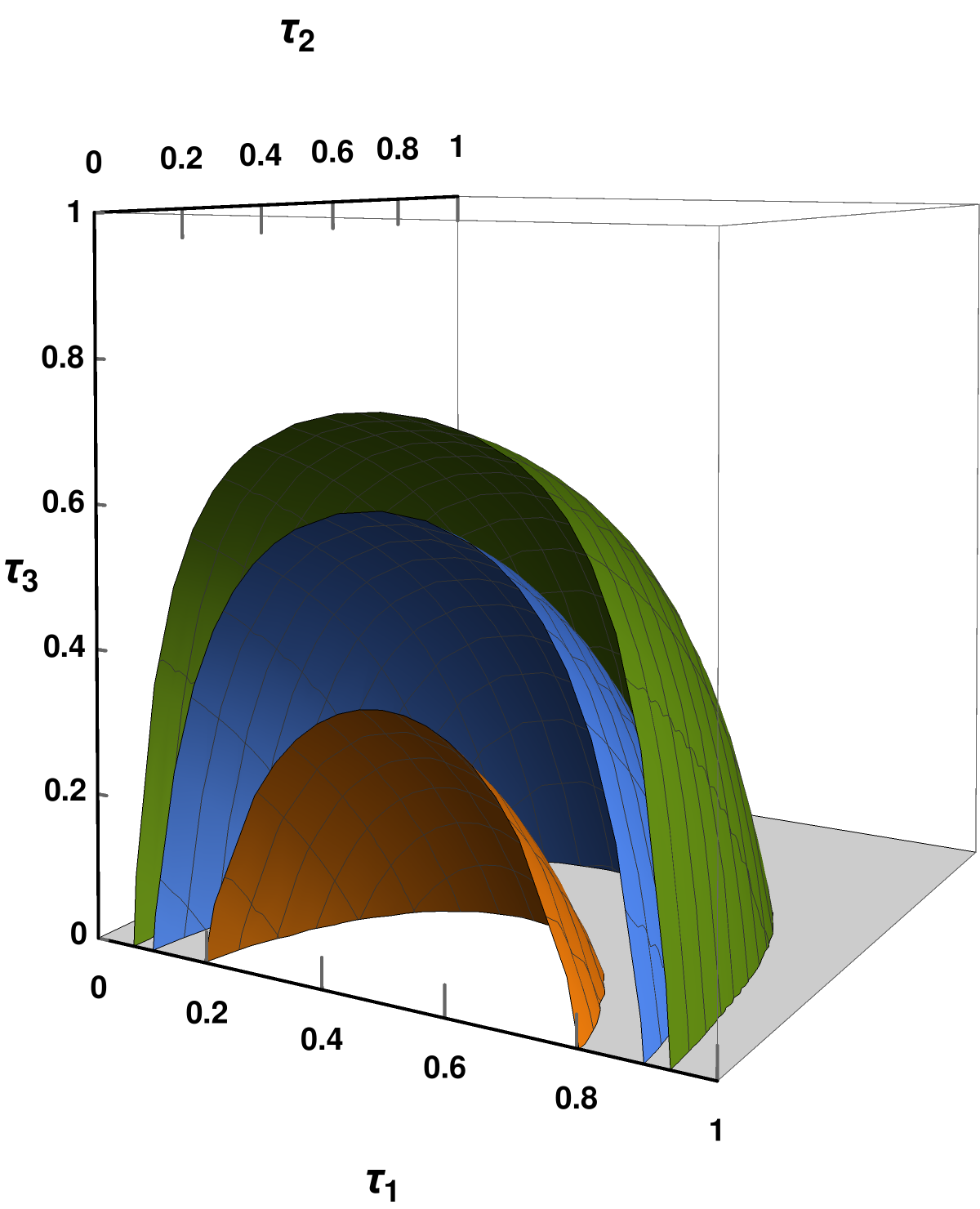}
    \caption{ 
    Quantum advantage (in bits) in the DC scheme involving three senders and a single receiver  in the state space characterized by $\tau_1$ (x-axis), $\tau_2$ (y-axis) and $\tau_3$ (z-axis). The orange, blue and green planes represent $\bar{N} = 15$, $\bar{N} = 20$ and $\bar{N} = 25$ respectively. Note that the volume enclosed by the surfaces represents states which can provide a quantum advantage, i.e., \(\Delta^{\mathcal{S}_1\mathcal{S}_2\mathcal{S}_3:\mathcal{R}} >0\). All the axes are dimensionless. }
    \label{fig:4mode_region}
\end{figure}


For the classical information transmission involving three senders and a single receiver without having any shared entangled state, the classical threshold reduces to 
\begin{equation}
    C_{cl}^{\mathcal{S}_1\mathcal{S}_2\mathcal{S}_3:\mathcal{R}} = 3(1 + \bar{N}/3) \ln (1 + \bar{N}/3) - 3 (\bar{N}/3) \ln (\bar{N}/3).
    \label{eq:4mode_c-cap}
\end{equation}
Analyzing 
\(\Delta^{\mathcal{S}_1\mathcal{S}_2\mathcal{S}_3:\mathcal{R}}\) in the \(\tau_1, \tau_2, \tau_3\) hyperplane, we observe that  the quantum protocol can outperform the classical one for a given $\bar{N}$ as depicted in  Fig. \ref{fig:4mode_region}. The volume of states having  quantum benefit increases with the increase of  \(\bar{N}\) as also seen in the case of shared three-mode states and it is bounded by the surface in the figure. Moreover, we notice that 
all such favorable states are centered around $\tau_1 = 1/2$, which indicates that such a configuration is well suited for the proposed CVDC protocol between three senders and a single receiver. Furthermore, for small  signal strength at the senders' end, states with small values of $\tau_2$ and $\tau_3$ are more helpful over the classical scheme compared to the states with high values of transmission coefficients of the beam splitters.



Like the three-mode entangled case,
the solution of the equation,   $C^{\mathcal{S}_1\mathcal{S}_2\mathcal{S}_3:\mathcal{R}}(\tau_1,\tau_2,\tau_3)=C^{\mathcal{S}_1\mathcal{S}_2\mathcal{S}_3:\mathcal{R}}_{cl}$ for $\bar{N}$ can give the threshold energy,  $\bar{N}^{\mathcal{S}_1\mathcal{S}_2\mathcal{S}_3:\mathcal{R}}_{th}(\tau_1,\tau_2,\tau_3)$ for the shared state comprising state parameters $\tau_1$, $\tau_2$ and $\tau_3$, above which \(\Delta^{\mathcal{S}_1 \mathcal{S}_2 \mathcal{S}_3: \mathcal{R}} >0 \). E.g.
$\bar{N}^{\mathcal{S}_1\mathcal{S}_2\mathcal{S}_3:\mathcal{R}}_{th}(1/2,1/2,1/2)=24.87$ and 
\begin{equation*}
     \bar{N}^{\mathcal{S}_1\mathcal{S}_2\mathcal{S}_3:\mathcal{R}}_{th}\big|_{\min\limits_{\{\tau_1,\tau_2,\tau_3\}}} = 11.45,
\end{equation*}
for the shared four-mode genuinely multimode entangled states.
In this situation, let us identify the range of state parameters, i.e., 
$\tau_1$, $\tau_2$ and $\tau_3$ for a given energy $\bar{N}$ so that the quantum advantage can be prevailed. They turn out to be 

\begin{widetext}
\begin{eqnarray}
\tau_1\big|_{\substack{\max\\\min}} =  0.5 \pm
\sqrt{\frac{9(\bar{N} (\bar{N}+3)+6)^2-\frac{4 \bar{N}^{-2 \bar{N}} (\bar{N}+3)^{2 \bar{N}+6}}{(\bar{N} (\bar{N}+3)+3)^2}}{36 \bar{N}^2 (\bar{N}+3)^2}},
\end{eqnarray}
while for given $\bar{N}$ and $\tau_1$,
\begin{eqnarray}
\tau_2\big|_{max}=\frac{\frac{(\bar{N}+3)^{2 \bar{N}+6} \bar{N}^{-2 \bar{N}}}{(\bar{N} (\bar{N}+3)+3)^2 \left(\bar{N} (\bar{N}+3) \left(\tau _1-1\right)-3\right)}+9 (\bar{N}+3) \bar{N} \tau _1+27}{9 \bar{N} (\bar{N}+3) \tau _1}.
\end{eqnarray}
When  $\bar{N}$, $\tau_1$ and $\tau_2$ are fixed, the third transmission coefficient  takes the form as
\begin{eqnarray}
\tau_3\big|_{max}=\frac{9 (\bar{N} (\bar{N}+3)+3)^2 \left(\bar{N} (\bar{N}+3) \left(\tau _1-1\right)-3\right) \left(\bar{N} (\bar{N}+3) \tau _1 \left(\tau _2-1\right)-3\right)-\bar{N}^{-2 \bar{N}} (\bar{N}+3)^{2 \bar{N}+6}}{9 \bar{N} (\bar{N}+3) (\bar{N} (\bar{N}+3)+3) \left(\bar{N} (\bar{N}+3) \tau _1 \left(\tau _2-1\right)-3\right) \left((\bar{N} (\bar{N}+3)+3) \left(\tau _1-1\right)-3 \tau _1 \tau _2\right)}.
\end{eqnarray}
\end{widetext}
Substituting $\bar{N}=3e^{r}\sinh{r}$ into Eqs. (\ref{eq:4mode_cap}) and (\ref{eq:4mode_c-cap}), we obtain $C^{\mathcal{S}_1\mathcal{S}_2\mathcal{S}_3:\mathcal{R}}\sim8r$ and $C^{\mathcal{S}_1\mathcal{S}_2\mathcal{S}_3:\mathcal{R}}_{cl}\sim6r$ respectively for large $r$. Therefore,  the ratio between quantum and classical protocols for three senders and a single receiver  becomes
\begin{equation}
   \frac{C^{\mathcal{S}_1\mathcal{S}_2\mathcal{S}_3:\mathcal{R}}}{C^{\mathcal{S}_1\mathcal{S}_2\mathcal{S}_3:\mathcal{R}}_{cl}} = \frac{4}{3} \;\;\;\;\;\;\text{(at large $r$)}.
   \label{eq:ratio_4}
\end{equation}

In this case, the break-even squeezing strength of the quantum protocol, given in Eq. \eqref{eq:4mode_opt_cap}, required to defeat the classical threshold with the DC capacity for a given senders' photon number $\bar{N}$ reads $r_{break-even}^{\mathcal{S}_1\mathcal{S}_2\mathcal{S}_3:\mathcal{R}} (\tau_1 = \tau_2 = \tau_3=1/2) = 1.433$ which is \(1.107\) for the three-mode case with \(\tau_1 = \tau_2 = 1/2\). It  implies that the squeezing strength required to obtain improvement in the mentioned quantum protocol     increases with the increase of the number of modes.
Thus for the three senders-one receiver scenario, there is a quantum advantage beyond $r_{break-even}^{\mathcal{S}_1\mathcal{S}_2\mathcal{S}_3:\mathcal{R}} (\tau_1, \tau_2, \tau_3)$.\\
\textcolor{black}{At this point, it can possibly be argued, that for $\mathcal{N}$ senders and a single receiver, the ratio between the capacities of the  quantum  and classical channels takes the  form,
\begin{equation}
    \frac{C^{\mathcal{S}_1\dots\mathcal{S}_\mathcal{N}:\mathcal{R}}}{C^{\mathcal{S}_1\dots\mathcal{S}_\mathcal{N}:\mathcal{R}}_{cl}} = \frac{\mathcal{N} + 1}{\mathcal{N}},
    \label{eq:ratio_N}
\end{equation}
at large $r$, where we have used Eqs. \eqref{eq:ratio_3} and \eqref{eq:ratio_4} to present this conjecture.}

\section{Conclusion}
\label{sec:conclu}

In quantum communication which includes both classical information transmission as well as quantum state transfer, shared entangled states are necessary to exhibit any quantum advantage. 
To transfer classical information, say two bits, the classical protocol where no shared entangled state is available requires four-dimensional objects for encoding while it reduces to a two-dimensional system with the help of  shared entangled states and hence the scheme is called  dense coding (DC). In finite dimensional systems, the capacity of dense coding for an arbitrary shared state is known when there are an arbitrary number of senders and a single or two receivers.      

For continuous variable (CV) systems, since the dimension of the systems involved is infinite, the DC capacity can only be  meaningful when it is 
obtained by fixing the amount of energy that can be sent from the sender to the receiver. Without this constraint, the capacity would simply diverge. Using this energy-constrained capacity, the quantum advantage in CVDC was demonstrated for a single sender and a single receiver scenario  \cite{Braunstein_PRA_2000}. 

In this work, we have gone beyond the single sender-receiver scenario, and have proposed a design for continuous variable DC networks with multiple senders and a single receiver. In particular,  
we have presented a possible blueprint of  the encoding as well as decoding strategies, computed the corresponding classical energy-constrained capacities of a quantum channel, and optimum classical threshold which can be achieved in absence of a shared entangled state. 
We have fixed the encoding strategies to be local displacement operations on the senders' side, while the decoding  involves the use of  beam splitters and the homodyne measurement of quadratures.

We have demonstrated the efficacy of the CVDC network involving  two as well as three senders and a single receiver  when the shared states are the  three- and four-mode states. In both cases, we have shown that the quantum protocol can give benefit over the classical one, thereby establishing the usefulness of  multimode entangled states as resources. With the increase of energy, we have found that the quantum advantage also got enhanced. Moreover, we  have computed the critical energy which is required for the successful implementation of  CVDC with an entangled resource. 

A practical communication technology demands the transfer of data among various nodes in a network. Hence the construction of the protocol presented here  may shed light to establish  a network for transmitting classical information involving multiple nodes using squeezed states of light  which can be implementable in laboratories. 

	\section{Acknowledgement}
	
	AP, RG, and ASD acknowledge the support from Interdisciplinary Cyber Physical Systems (ICPS) program of the Department of Science and Technology (DST), India, Grant No.: DST/ICPS/QuST/Theme- 1/2019/23. TD acknowledges support by the Foundation for Polish Science (IRAP project, ICTQT, contract no. MAB/2018/5, co-financed by EU within Smart Growth Operational Programme. This work has been partly supported by the Hong Kong Research Grant Council (RGC) through grant 17300918.

\bibliographystyle{apsrev4-1}
	\bibliography{reference}

\begin{thebibliography}{56}%
\makeatletter
\providecommand \@ifxundefined [1]{%
 \@ifx{#1\undefined}
}%
\providecommand \@ifnum [1]{%
 \ifnum #1\expandafter \@firstoftwo
 \else \expandafter \@secondoftwo
 \fi
}%
\providecommand \@ifx [1]{%
 \ifx #1\expandafter \@firstoftwo
 \else \expandafter \@secondoftwo
 \fi
}%
\providecommand \natexlab [1]{#1}%
\providecommand \enquote  [1]{``#1''}%
\providecommand \bibnamefont  [1]{#1}%
\providecommand \bibfnamefont [1]{#1}%
\providecommand \citenamefont [1]{#1}%
\providecommand \href@noop [0]{\@secondoftwo}%
\providecommand \href [0]{\begingroup \@sanitize@url \@href}%
\providecommand \@href[1]{\@@startlink{#1}\@@href}%
\providecommand \@@href[1]{\endgroup#1\@@endlink}%
\providecommand \@sanitize@url [0]{\catcode `\\12\catcode `\$12\catcode
  `\&12\catcode `\#12\catcode `\^12\catcode `\_12\catcode `\%12\relax}%
\providecommand \@@startlink[1]{}%
\providecommand \@@endlink[0]{}%
\providecommand \url  [0]{\begingroup\@sanitize@url \@url }%
\providecommand \@url [1]{\endgroup\@href {#1}{\urlprefix }}%
\providecommand \urlprefix  [0]{URL }%
\providecommand \Eprint [0]{\href }%
\providecommand \doibase [0]{http://dx.doi.org/}%
\providecommand \selectlanguage [0]{\@gobble}%
\providecommand \bibinfo  [0]{\@secondoftwo}%
\providecommand \bibfield  [0]{\@secondoftwo}%
\providecommand \translation [1]{[#1]}%
\providecommand \BibitemOpen [0]{}%
\providecommand \bibitemStop [0]{}%
\providecommand \bibitemNoStop [0]{.\EOS\space}%
\providecommand \EOS [0]{\spacefactor3000\relax}%
\providecommand \BibitemShut  [1]{\csname bibitem#1\endcsname}%
\let\auto@bib@innerbib\@empty
\bibitem [{\citenamefont {Bennett}\ and\ \citenamefont
  {Brassard}(2014)}]{BB84}%
  \BibitemOpen
  \bibfield  {author} {\bibinfo {author} {\bibfnamefont {C.~H.}\ \bibnamefont
  {Bennett}}\ and\ \bibinfo {author} {\bibfnamefont {G.}~\bibnamefont
  {Brassard}},\ }\href {\doibase https://doi.org/10.1016/j.tcs.2014.05.025}
  {\bibfield  {journal} {\bibinfo  {journal} {Theoretical Computer Science}\
  }\textbf {\bibinfo {volume} {560}},\ \bibinfo {pages} {7} (\bibinfo {year}
  {2014})}\BibitemShut {NoStop}%
\bibitem [{\citenamefont {Ekert}(1991)}]{EkertCrypto}%
  \BibitemOpen
  \bibfield  {author} {\bibinfo {author} {\bibfnamefont {A.~K.}\ \bibnamefont
  {Ekert}},\ }\href {\doibase 10.1103/PhysRevLett.67.661} {\bibfield  {journal}
  {\bibinfo  {journal} {Phys. Rev. Lett.}\ }\textbf {\bibinfo {volume} {67}},\
  \bibinfo {pages} {661} (\bibinfo {year} {1991})}\BibitemShut {NoStop}%
\bibitem [{\citenamefont {Jennewein}\ \emph {et~al.}(2000)\citenamefont
  {Jennewein}, \citenamefont {Simon}, \citenamefont {Weihs}, \citenamefont
  {Weinfurter},\ and\ \citenamefont {Zeilinger}}]{Crypto3}%
  \BibitemOpen
  \bibfield  {author} {\bibinfo {author} {\bibfnamefont {T.}~\bibnamefont
  {Jennewein}}, \bibinfo {author} {\bibfnamefont {C.}~\bibnamefont {Simon}},
  \bibinfo {author} {\bibfnamefont {G.}~\bibnamefont {Weihs}}, \bibinfo
  {author} {\bibfnamefont {H.}~\bibnamefont {Weinfurter}}, \ and\ \bibinfo
  {author} {\bibfnamefont {A.}~\bibnamefont {Zeilinger}},\ }\href {\doibase
  10.1103/PhysRevLett.84.4729} {\bibfield  {journal} {\bibinfo  {journal}
  {Phys. Rev. Lett.}\ }\textbf {\bibinfo {volume} {84}},\ \bibinfo {pages}
  {4729} (\bibinfo {year} {2000})}\BibitemShut {NoStop}%
\bibitem [{\citenamefont {Gisin}\ \emph {et~al.}(2002)\citenamefont {Gisin},
  \citenamefont {Ribordy}, \citenamefont {Tittel},\ and\ \citenamefont
  {Zbinden}}]{Crypto4}%
  \BibitemOpen
  \bibfield  {author} {\bibinfo {author} {\bibfnamefont {N.}~\bibnamefont
  {Gisin}}, \bibinfo {author} {\bibfnamefont {G.}~\bibnamefont {Ribordy}},
  \bibinfo {author} {\bibfnamefont {W.}~\bibnamefont {Tittel}}, \ and\ \bibinfo
  {author} {\bibfnamefont {H.}~\bibnamefont {Zbinden}},\ }\href {\doibase
  10.1103/RevModPhys.74.145} {\bibfield  {journal} {\bibinfo  {journal} {Rev.
  Mod. Phys.}\ }\textbf {\bibinfo {volume} {74}},\ \bibinfo {pages} {145}
  (\bibinfo {year} {2002})}\BibitemShut {NoStop}%
\bibitem [{\citenamefont {Vazirani}\ and\ \citenamefont
  {Vidick}(2014)}]{Vazirani_PRL_2014}%
  \BibitemOpen
  \bibfield  {author} {\bibinfo {author} {\bibfnamefont {U.}~\bibnamefont
  {Vazirani}}\ and\ \bibinfo {author} {\bibfnamefont {T.}~\bibnamefont
  {Vidick}},\ }\href {\doibase 10.1103/PhysRevLett.113.140501} {\bibfield
  {journal} {\bibinfo  {journal} {Phys. Rev. Lett.}\ }\textbf {\bibinfo
  {volume} {113}},\ \bibinfo {pages} {140501} (\bibinfo {year}
  {2014})}\BibitemShut {NoStop}%
\bibitem [{\citenamefont {Mayers}\ and\ \citenamefont
  {Yao}(1998)}]{Mayers_arxiv_1998}%
  \BibitemOpen
  \bibfield  {author} {\bibinfo {author} {\bibfnamefont {D.}~\bibnamefont
  {Mayers}}\ and\ \bibinfo {author} {\bibfnamefont {A.}~\bibnamefont {Yao}},\
  }\bibfield  {booktitle} {\emph {\bibinfo {booktitle} {Proceedings 39th Annual
  Symposium on Foundations of Computer Science (Cat. No.98CB36280)}},\ }\href
  {\doibase 10.1109/SFCS.1998.743501} {\ ,\ \bibinfo {pages} {503} (\bibinfo
  {year} {1998})}\BibitemShut {NoStop}%
\bibitem [{\citenamefont {Miller}\ and\ \citenamefont
  {Shi}(2016)}]{Miller_JACM_2016}%
  \BibitemOpen
  \bibfield  {author} {\bibinfo {author} {\bibfnamefont {C.~A.}\ \bibnamefont
  {Miller}}\ and\ \bibinfo {author} {\bibfnamefont {Y.}~\bibnamefont {Shi}},\
  }\href {\doibase 10.1145/2885493} {\bibfield  {journal} {\bibinfo  {journal}
  {J. ACM}\ }\textbf {\bibinfo {volume} {63}} (\bibinfo {year} {2016}),\
  10.1145/2885493}\BibitemShut {NoStop}%
\bibitem [{\citenamefont {Bennett}\ and\ \citenamefont
  {Wiesner}(1992)}]{bennettwiesner}%
  \BibitemOpen
  \bibfield  {author} {\bibinfo {author} {\bibfnamefont {C.~H.}\ \bibnamefont
  {Bennett}}\ and\ \bibinfo {author} {\bibfnamefont {S.~J.}\ \bibnamefont
  {Wiesner}},\ }\href {\doibase 10.1103/PhysRevLett.69.2881} {\bibfield
  {journal} {\bibinfo  {journal} {Phys. Rev. Lett.}\ }\textbf {\bibinfo
  {volume} {69}},\ \bibinfo {pages} {2881} (\bibinfo {year}
  {1992})}\BibitemShut {NoStop}%
\bibitem [{\citenamefont {Sen(De)}\ and\ \citenamefont
  {Sen}(2010)}]{AditiComm}%
  \BibitemOpen
  \bibfield  {author} {\bibinfo {author} {\bibfnamefont {A.}~\bibnamefont
  {Sen(De)}}\ and\ \bibinfo {author} {\bibfnamefont {U.}~\bibnamefont {Sen}},\
  }\href {\doibase .} {\bibfield  {journal} {\bibinfo  {journal} {Physics
  News}\ }\textbf {\bibinfo {volume} {40}},\ \bibinfo {pages} {17} (\bibinfo
  {year} {2010})}\BibitemShut {NoStop}%
\bibitem [{\citenamefont {Gisin}\ and\ \citenamefont {Thew}(2007)}]{GisinComm}%
  \BibitemOpen
  \bibfield  {author} {\bibinfo {author} {\bibfnamefont {N.}~\bibnamefont
  {Gisin}}\ and\ \bibinfo {author} {\bibfnamefont {R.}~\bibnamefont {Thew}},\
  }\href {\doibase https://doi.org/10.1038/nphoton.2007.22} {\bibfield
  {journal} {\bibinfo  {journal} {Nat. Photon.}\ }\textbf {\bibinfo {volume}
  {1}},\ \bibinfo {pages} {165} (\bibinfo {year} {2007})}\BibitemShut {NoStop}%
\bibitem [{\citenamefont {Demkowicz-Dobrza\ifmmode~\acute{n}\else
  \'{n}\fi{}ski}\ \emph {et~al.}(2009)\citenamefont
  {Demkowicz-Dobrza\ifmmode~\acute{n}\else \'{n}\fi{}ski}, \citenamefont
  {Sen(De)}, \citenamefont {Sen},\ and\ \citenamefont
  {Lewenstein}}]{AditiComm2}%
  \BibitemOpen
  \bibfield  {author} {\bibinfo {author} {\bibfnamefont {R.}~\bibnamefont
  {Demkowicz-Dobrza\ifmmode~\acute{n}\else \'{n}\fi{}ski}}, \bibinfo {author}
  {\bibfnamefont {A.}~\bibnamefont {Sen(De)}}, \bibinfo {author} {\bibfnamefont
  {U.}~\bibnamefont {Sen}}, \ and\ \bibinfo {author} {\bibfnamefont
  {M.}~\bibnamefont {Lewenstein}},\ }\href {\doibase
  10.1103/PhysRevA.80.012311} {\bibfield  {journal} {\bibinfo  {journal} {Phys.
  Rev. A}\ }\textbf {\bibinfo {volume} {80}},\ \bibinfo {pages} {012311}
  (\bibinfo {year} {2009})}\BibitemShut {NoStop}%
\bibitem [{\citenamefont {Sen(De)}\ \emph {et~al.}(2003)\citenamefont
  {Sen(De)}, \citenamefont {Sen},\ and\ \citenamefont {\ifmmode~\dot{Z}\else
  \.{Z}\fi{}ukowski}}]{AditiComm3}%
  \BibitemOpen
  \bibfield  {author} {\bibinfo {author} {\bibfnamefont {A.}~\bibnamefont
  {Sen(De)}}, \bibinfo {author} {\bibfnamefont {U.}~\bibnamefont {Sen}}, \ and\
  \bibinfo {author} {\bibfnamefont {M.}~\bibnamefont {\ifmmode~\dot{Z}\else
  \.{Z}\fi{}ukowski}},\ }\href {\doibase 10.1103/PhysRevA.68.032309} {\bibfield
   {journal} {\bibinfo  {journal} {Phys. Rev. A}\ }\textbf {\bibinfo {volume}
  {68}},\ \bibinfo {pages} {032309} (\bibinfo {year} {2003})}\BibitemShut
  {NoStop}%
\bibitem [{\citenamefont {Bennett}\ \emph {et~al.}(1993)\citenamefont
  {Bennett}, \citenamefont {Brassard}, \citenamefont {Cr\'epeau}, \citenamefont
  {Jozsa}, \citenamefont {Peres},\ and\ \citenamefont {Wootters}}]{BBCJPW}%
  \BibitemOpen
  \bibfield  {author} {\bibinfo {author} {\bibfnamefont {C.~H.}\ \bibnamefont
  {Bennett}}, \bibinfo {author} {\bibfnamefont {G.}~\bibnamefont {Brassard}},
  \bibinfo {author} {\bibfnamefont {C.}~\bibnamefont {Cr\'epeau}}, \bibinfo
  {author} {\bibfnamefont {R.}~\bibnamefont {Jozsa}}, \bibinfo {author}
  {\bibfnamefont {A.}~\bibnamefont {Peres}}, \ and\ \bibinfo {author}
  {\bibfnamefont {W.~K.}\ \bibnamefont {Wootters}},\ }\href {\doibase
  10.1103/PhysRevLett.70.1895} {\bibfield  {journal} {\bibinfo  {journal}
  {Phys. Rev. Lett.}\ }\textbf {\bibinfo {volume} {70}},\ \bibinfo {pages}
  {1895} (\bibinfo {year} {1993})}\BibitemShut {NoStop}%
\bibitem [{\citenamefont {Vaidman}(1994)}]{Vaidman94}%
  \BibitemOpen
  \bibfield  {author} {\bibinfo {author} {\bibfnamefont {L.}~\bibnamefont
  {Vaidman}},\ }\href {\doibase 10.1103/PhysRevA.49.1473} {\bibfield  {journal}
  {\bibinfo  {journal} {Phys. Rev. A}\ }\textbf {\bibinfo {volume} {49}},\
  \bibinfo {pages} {1473} (\bibinfo {year} {1994})}\BibitemShut {NoStop}%
\bibitem [{\citenamefont {Braunstein}\ and\ \citenamefont
  {Kimble}(1998)}]{Braunstein_PRL_1998}%
  \BibitemOpen
  \bibfield  {author} {\bibinfo {author} {\bibfnamefont {S.~L.}\ \bibnamefont
  {Braunstein}}\ and\ \bibinfo {author} {\bibfnamefont {H.~J.}\ \bibnamefont
  {Kimble}},\ }\href {\doibase 10.1103/PhysRevLett.80.869} {\bibfield
  {journal} {\bibinfo  {journal} {Phys. Rev. Lett.}\ }\textbf {\bibinfo
  {volume} {80}},\ \bibinfo {pages} {869} (\bibinfo {year} {1998})}\BibitemShut
  {NoStop}%
\bibitem [{\citenamefont {Raussendorf}\ and\ \citenamefont
  {Briegel}(2001)}]{Raussendorf_PRL_2001}%
  \BibitemOpen
  \bibfield  {author} {\bibinfo {author} {\bibfnamefont {R.}~\bibnamefont
  {Raussendorf}}\ and\ \bibinfo {author} {\bibfnamefont {H.~J.}\ \bibnamefont
  {Briegel}},\ }\href {\doibase 10.1103/PhysRevLett.86.5188} {\bibfield
  {journal} {\bibinfo  {journal} {Phys. Rev. Lett.}\ }\textbf {\bibinfo
  {volume} {86}},\ \bibinfo {pages} {5188} (\bibinfo {year}
  {2001})}\BibitemShut {NoStop}%
\bibitem [{\citenamefont {Briegel}\ and\ \citenamefont
  {Raussendorf}(2001)}]{Briegel_PRL_2001}%
  \BibitemOpen
  \bibfield  {author} {\bibinfo {author} {\bibfnamefont {H.~J.}\ \bibnamefont
  {Briegel}}\ and\ \bibinfo {author} {\bibfnamefont {R.}~\bibnamefont
  {Raussendorf}},\ }\href {\doibase 10.1103/PhysRevLett.86.910} {\bibfield
  {journal} {\bibinfo  {journal} {Phys. Rev. Lett.}\ }\textbf {\bibinfo
  {volume} {86}},\ \bibinfo {pages} {910} (\bibinfo {year} {2001})}\BibitemShut
  {NoStop}%
\bibitem [{\citenamefont {Raussendorf}\ \emph {et~al.}(2003)\citenamefont
  {Raussendorf}, \citenamefont {Browne},\ and\ \citenamefont
  {Briegel}}]{Raussendorf_PRA_2003}%
  \BibitemOpen
  \bibfield  {author} {\bibinfo {author} {\bibfnamefont {R.}~\bibnamefont
  {Raussendorf}}, \bibinfo {author} {\bibfnamefont {D.~E.}\ \bibnamefont
  {Browne}}, \ and\ \bibinfo {author} {\bibfnamefont {H.~J.}\ \bibnamefont
  {Briegel}},\ }\href {\doibase 10.1103/PhysRevA.68.022312} {\bibfield
  {journal} {\bibinfo  {journal} {Phys. Rev. A}\ }\textbf {\bibinfo {volume}
  {68}},\ \bibinfo {pages} {022312} (\bibinfo {year} {2003})}\BibitemShut
  {NoStop}%
\bibitem [{\citenamefont {Walther}\ \emph {et~al.}(2005)\citenamefont
  {Walther}, \citenamefont {Resch}, \citenamefont {Rudolph}, \citenamefont
  {Schenck}, \citenamefont {Weinfurter}, \citenamefont {Vedral}, \citenamefont
  {Aspelmeyer},\ and\ \citenamefont {Zeilinger}}]{Walther_Nature_2005}%
  \BibitemOpen
  \bibfield  {author} {\bibinfo {author} {\bibfnamefont {P.}~\bibnamefont
  {Walther}}, \bibinfo {author} {\bibfnamefont {K.~J.}\ \bibnamefont {Resch}},
  \bibinfo {author} {\bibfnamefont {T.}~\bibnamefont {Rudolph}}, \bibinfo
  {author} {\bibfnamefont {E.}~\bibnamefont {Schenck}}, \bibinfo {author}
  {\bibfnamefont {H.}~\bibnamefont {Weinfurter}}, \bibinfo {author}
  {\bibfnamefont {V.}~\bibnamefont {Vedral}}, \bibinfo {author} {\bibfnamefont
  {M.}~\bibnamefont {Aspelmeyer}}, \ and\ \bibinfo {author} {\bibfnamefont
  {A.}~\bibnamefont {Zeilinger}},\ }\href {\doibase 10.1038/nature03347}
  {\bibfield  {journal} {\bibinfo  {journal} {Nature}\ }\textbf {\bibinfo
  {volume} {434}},\ \bibinfo {pages} {169} (\bibinfo {year}
  {2005})}\BibitemShut {NoStop}%
\bibitem [{\citenamefont {Raussendorf}\ \emph {et~al.}(2007)\citenamefont
  {Raussendorf}, \citenamefont {Harrington},\ and\ \citenamefont
  {Goyal}}]{Raussendorf_NJP_2007}%
  \BibitemOpen
  \bibfield  {author} {\bibinfo {author} {\bibfnamefont {R.}~\bibnamefont
  {Raussendorf}}, \bibinfo {author} {\bibfnamefont {J.}~\bibnamefont
  {Harrington}}, \ and\ \bibinfo {author} {\bibfnamefont {K.}~\bibnamefont
  {Goyal}},\ }\href {\doibase 10.1088/1367-2630/9/6/199} {\bibfield  {journal}
  {\bibinfo  {journal} {New Journal of Physics}\ }\textbf {\bibinfo {volume}
  {9}},\ \bibinfo {pages} {199} (\bibinfo {year} {2007})}\BibitemShut {NoStop}%
\bibitem [{\citenamefont {Raussendorf}\ and\ \citenamefont
  {Harrington}(2007)}]{Raussendorf_PRL_2007}%
  \BibitemOpen
  \bibfield  {author} {\bibinfo {author} {\bibfnamefont {R.}~\bibnamefont
  {Raussendorf}}\ and\ \bibinfo {author} {\bibfnamefont {J.}~\bibnamefont
  {Harrington}},\ }\href {\doibase 10.1103/PhysRevLett.98.190504} {\bibfield
  {journal} {\bibinfo  {journal} {Phys. Rev. Lett.}\ }\textbf {\bibinfo
  {volume} {98}},\ \bibinfo {pages} {190504} (\bibinfo {year}
  {2007})}\BibitemShut {NoStop}%
\bibitem [{\citenamefont {Verstraete}\ \emph {et~al.}(2009)\citenamefont
  {Verstraete}, \citenamefont {Wolf},\ and\ \citenamefont
  {Ignacio}}]{Verstraete_Nature_2009}%
  \BibitemOpen
  \bibfield  {author} {\bibinfo {author} {\bibfnamefont {F.}~\bibnamefont
  {Verstraete}}, \bibinfo {author} {\bibfnamefont {M.~M.}\ \bibnamefont
  {Wolf}}, \ and\ \bibinfo {author} {\bibfnamefont {C.~J.}\ \bibnamefont
  {Ignacio}},\ }\href {\doibase 10.1038/nphys1342} {\bibfield  {journal}
  {\bibinfo  {journal} {Nature Physics}\ }\textbf {\bibinfo {volume} {5}},\
  \bibinfo {pages} {633} (\bibinfo {year} {2009})}\BibitemShut {NoStop}%
\bibitem [{\citenamefont {Ma}\ \emph {et~al.}(2016)\citenamefont {Ma},
  \citenamefont {Yuan}, \citenamefont {Cao}, \citenamefont {Qi},\ and\
  \citenamefont {Zhang}}]{Ma_NPJ_2016}%
  \BibitemOpen
  \bibfield  {author} {\bibinfo {author} {\bibfnamefont {X.}~\bibnamefont
  {Ma}}, \bibinfo {author} {\bibfnamefont {X.}~\bibnamefont {Yuan}}, \bibinfo
  {author} {\bibfnamefont {Z.}~\bibnamefont {Cao}}, \bibinfo {author}
  {\bibfnamefont {B.}~\bibnamefont {Qi}}, \ and\ \bibinfo {author}
  {\bibfnamefont {Z.}~\bibnamefont {Zhang}},\ }\href {\doibase
  10.1038/npjqi.2016.21} {\bibfield  {journal} {\bibinfo  {journal} {npj
  Quantum Information}\ }\textbf {\bibinfo {volume} {2}},\ \bibinfo {pages}
  {16021} (\bibinfo {year} {2016})}\BibitemShut {NoStop}%
\bibitem [{\citenamefont {Kollmitzer}\ \emph {et~al.}(1997)\citenamefont
  {Kollmitzer}, \citenamefont {Schaur}, \citenamefont {Rass},\ and\
  \citenamefont {Rainer}}]{Kollmitzer_Springer_2020}%
  \BibitemOpen
  \bibfield  {author} {\bibinfo {author} {\bibfnamefont {C.}~\bibnamefont
  {Kollmitzer}}, \bibinfo {author} {\bibfnamefont {S.}~\bibnamefont {Schaur}},
  \bibinfo {author} {\bibfnamefont {S.}~\bibnamefont {Rass}}, \ and\ \bibinfo
  {author} {\bibfnamefont {B.}~\bibnamefont {Rainer}},\ }\href {\doibase
  https://doi.org/10.1007/978-3-319-72596-3} {\emph {\bibinfo {title} {Quantum
  Random Number Generation}}}\ (\bibinfo  {publisher} {Springer, Cham},\
  \bibinfo {year} {1997})\BibitemShut {NoStop}%
\bibitem [{\citenamefont {Bruß}\ \emph {et~al.}(2014)\citenamefont {Bruß},
  \citenamefont {DAriano}, \citenamefont {Lewenstein}, \citenamefont
  {Macchiavello}, \citenamefont {Sen(De)},\ and\ \citenamefont {Sen}}]{Bruss}%
  \BibitemOpen
  \bibfield  {author} {\bibinfo {author} {\bibfnamefont {D.}~\bibnamefont
  {Bruß}}, \bibinfo {author} {\bibfnamefont {G.~M.}\ \bibnamefont {DAriano}},
  \bibinfo {author} {\bibfnamefont {M.}~\bibnamefont {Lewenstein}}, \bibinfo
  {author} {\bibfnamefont {C.}~\bibnamefont {Macchiavello}}, \bibinfo {author}
  {\bibfnamefont {A.}~\bibnamefont {Sen(De)}}, \ and\ \bibinfo {author}
  {\bibfnamefont {U.}~\bibnamefont {Sen}},\ }\href {\doibase
  https://doi.org/10.1103/PhysRevLett.93.210501} {\bibfield  {journal}
  {\bibinfo  {journal} {Phys. Rev. Lett.}\ }\textbf {\bibinfo {volume} {93}},\
  \bibinfo {pages} {210501} (\bibinfo {year} {2014})}\BibitemShut {NoStop}%
\bibitem [{\citenamefont {Bru\ss{}}\ \emph {et~al.}(2006)\citenamefont
  {Bru\ss{}}, \citenamefont {Lewenstein}, \citenamefont {Sen(De)},
  \citenamefont {Sen}, \citenamefont {D'Ariano},\ and\ \citenamefont
  {Macchiavello}}]{DCCamader}%
  \BibitemOpen
  \bibfield  {author} {\bibinfo {author} {\bibfnamefont {D.}~\bibnamefont
  {Bru\ss{}}}, \bibinfo {author} {\bibfnamefont {M.}~\bibnamefont
  {Lewenstein}}, \bibinfo {author} {\bibfnamefont {A.}~\bibnamefont {Sen(De)}},
  \bibinfo {author} {\bibfnamefont {U.}~\bibnamefont {Sen}}, \bibinfo {author}
  {\bibfnamefont {G.~M.}\ \bibnamefont {D'Ariano}}, \ and\ \bibinfo {author}
  {\bibfnamefont {C.}~\bibnamefont {Macchiavello}},\ }\href {\doibase
  https://doi.org/10.1142/S0219749906001888} {\bibfield  {journal} {\bibinfo
  {journal} {International Journal of Quantum Information}\ }\textbf {\bibinfo
  {volume} {4}},\ \bibinfo {pages} {415} (\bibinfo {year} {2006})}\BibitemShut
  {NoStop}%
\bibitem [{\citenamefont {Das}\ \emph {et~al.}(2015)\citenamefont {Das},
  \citenamefont {Prabhu}, \citenamefont {Sen(De)},\ and\ \citenamefont
  {Sen}}]{DCCTamo1}%
  \BibitemOpen
  \bibfield  {author} {\bibinfo {author} {\bibfnamefont {T.}~\bibnamefont
  {Das}}, \bibinfo {author} {\bibfnamefont {R.}~\bibnamefont {Prabhu}},
  \bibinfo {author} {\bibfnamefont {A.}~\bibnamefont {Sen(De)}}, \ and\
  \bibinfo {author} {\bibfnamefont {U.}~\bibnamefont {Sen}},\ }\href {\doibase
  10.1103/PhysRevA.92.052330} {\bibfield  {journal} {\bibinfo  {journal} {Phys.
  Rev. A}\ }\textbf {\bibinfo {volume} {92}},\ \bibinfo {pages} {052330}
  (\bibinfo {year} {2015})}\BibitemShut {NoStop}%
\bibitem [{\citenamefont {Das}\ \emph {et~al.}(2014)\citenamefont {Das},
  \citenamefont {Prabhu}, \citenamefont {Sen(De)},\ and\ \citenamefont
  {Sen}}]{Tamoghna}%
  \BibitemOpen
  \bibfield  {author} {\bibinfo {author} {\bibfnamefont {T.}~\bibnamefont
  {Das}}, \bibinfo {author} {\bibfnamefont {R.}~\bibnamefont {Prabhu}},
  \bibinfo {author} {\bibfnamefont {A.}~\bibnamefont {Sen(De)}}, \ and\
  \bibinfo {author} {\bibfnamefont {U.}~\bibnamefont {Sen}},\ }\href {\doibase
  10.1103/PhysRevA.90.022319} {\bibfield  {journal} {\bibinfo  {journal} {Phys.
  Rev. A}\ }\textbf {\bibinfo {volume} {90}},\ \bibinfo {pages} {022319}
  (\bibinfo {year} {2014})}\BibitemShut {NoStop}%
\bibitem [{\citenamefont {Prabhu}\ \emph {et~al.}(2013)\citenamefont {Prabhu},
  \citenamefont {Pati}, \citenamefont {Sen~(De)},\ and\ \citenamefont
  {Sen}}]{Prabhu}%
  \BibitemOpen
  \bibfield  {author} {\bibinfo {author} {\bibfnamefont {R.}~\bibnamefont
  {Prabhu}}, \bibinfo {author} {\bibfnamefont {A.~K.}\ \bibnamefont {Pati}},
  \bibinfo {author} {\bibfnamefont {A.}~\bibnamefont {Sen~(De)}}, \ and\
  \bibinfo {author} {\bibfnamefont {U.}~\bibnamefont {Sen}},\ }\href {\doibase
  10.1103/PhysRevA.87.052319} {\bibfield  {journal} {\bibinfo  {journal} {Phys.
  Rev. A}\ }\textbf {\bibinfo {volume} {87}},\ \bibinfo {pages} {052319}
  (\bibinfo {year} {2013})}\BibitemShut {NoStop}%
\bibitem [{\citenamefont {Lee}\ \emph {et~al.}(2014)\citenamefont {Lee},
  \citenamefont {Ji}, \citenamefont {Park},\ and\ \citenamefont
  {Nha}}]{Lee_PRA_2014}%
  \BibitemOpen
  \bibfield  {author} {\bibinfo {author} {\bibfnamefont {J.}~\bibnamefont
  {Lee}}, \bibinfo {author} {\bibfnamefont {S.-W.}\ \bibnamefont {Ji}},
  \bibinfo {author} {\bibfnamefont {J.}~\bibnamefont {Park}}, \ and\ \bibinfo
  {author} {\bibfnamefont {H.}~\bibnamefont {Nha}},\ }\href {\doibase
  10.1103/PhysRevA.90.022301} {\bibfield  {journal} {\bibinfo  {journal} {Phys.
  Rev. A}\ }\textbf {\bibinfo {volume} {90}},\ \bibinfo {pages} {022301}
  (\bibinfo {year} {2014})}\BibitemShut {NoStop}%
\bibitem [{\citenamefont {Czekaj}\ \emph {et~al.}(2010)\citenamefont {Czekaj},
  \citenamefont {Korbicz}, \citenamefont {Chhajlany},\ and\ \citenamefont
  {Horodecki}}]{Czekaj_PRA_2010}%
  \BibitemOpen
  \bibfield  {author} {\bibinfo {author} {\bibfnamefont {L.}~\bibnamefont
  {Czekaj}}, \bibinfo {author} {\bibfnamefont {J.~K.}\ \bibnamefont {Korbicz}},
  \bibinfo {author} {\bibfnamefont {R.~W.}\ \bibnamefont {Chhajlany}}, \ and\
  \bibinfo {author} {\bibfnamefont {P.}~\bibnamefont {Horodecki}},\ }\href
  {\doibase 10.1103/PhysRevA.82.020302} {\bibfield  {journal} {\bibinfo
  {journal} {Phys. Rev. A}\ }\textbf {\bibinfo {volume} {82}},\ \bibinfo
  {pages} {020302} (\bibinfo {year} {2010})}\BibitemShut {NoStop}%
\bibitem [{\citenamefont {Bostr\"om}\ and\ \citenamefont
  {Felbinger}(2002)}]{Bostrom_PRL_2002}%
  \BibitemOpen
  \bibfield  {author} {\bibinfo {author} {\bibfnamefont {K.}~\bibnamefont
  {Bostr\"om}}\ and\ \bibinfo {author} {\bibfnamefont {T.}~\bibnamefont
  {Felbinger}},\ }\href {\doibase 10.1103/PhysRevLett.89.187902} {\bibfield
  {journal} {\bibinfo  {journal} {Phys. Rev. Lett.}\ }\textbf {\bibinfo
  {volume} {89}},\ \bibinfo {pages} {187902} (\bibinfo {year}
  {2002})}\BibitemShut {NoStop}%
\bibitem [{\citenamefont {Beaudry}\ \emph {et~al.}(2013)\citenamefont
  {Beaudry}, \citenamefont {Lucamarini}, \citenamefont {Mancini},\ and\
  \citenamefont {Renner}}]{Beaudry_PRA_2013}%
  \BibitemOpen
  \bibfield  {author} {\bibinfo {author} {\bibfnamefont {N.~J.}\ \bibnamefont
  {Beaudry}}, \bibinfo {author} {\bibfnamefont {M.}~\bibnamefont {Lucamarini}},
  \bibinfo {author} {\bibfnamefont {S.}~\bibnamefont {Mancini}}, \ and\
  \bibinfo {author} {\bibfnamefont {R.}~\bibnamefont {Renner}},\ }\href
  {\doibase 10.1103/PhysRevA.88.062302} {\bibfield  {journal} {\bibinfo
  {journal} {Phys. Rev. A}\ }\textbf {\bibinfo {volume} {88}},\ \bibinfo
  {pages} {062302} (\bibinfo {year} {2013})}\BibitemShut {NoStop}%
\bibitem [{\citenamefont {Das}\ \emph {et~al.}(2021)\citenamefont {Das},
  \citenamefont {Horodecki},\ and\ \citenamefont {Pisarczyk}}]{Das_arXiv_2021}%
  \BibitemOpen
  \bibfield  {author} {\bibinfo {author} {\bibfnamefont {T.}~\bibnamefont
  {Das}}, \bibinfo {author} {\bibfnamefont {K.}~\bibnamefont {Horodecki}}, \
  and\ \bibinfo {author} {\bibfnamefont {R.}~\bibnamefont {Pisarczyk}},\ }\href
  {\doibase 10.48550/ARXIV.2106.13310} {\  (\bibinfo {year} {2021}),\
  10.48550/ARXIV.2106.13310}\BibitemShut {NoStop}%
\bibitem [{\citenamefont {Weinfurter}(1994)}]{Weinfurter_1994}%
  \BibitemOpen
  \bibfield  {author} {\bibinfo {author} {\bibfnamefont {H.}~\bibnamefont
  {Weinfurter}},\ }\href {\doibase 10.1209/0295-5075/25/8/001} {\bibfield
  {journal} {\bibinfo  {journal} {Europhysics Letters ({EPL})}\ }\textbf
  {\bibinfo {volume} {25}},\ \bibinfo {pages} {559} (\bibinfo {year}
  {1994})}\BibitemShut {NoStop}%
\bibitem [{\citenamefont {Mattle}\ \emph {et~al.}(1996)\citenamefont {Mattle},
  \citenamefont {Weinfurter}, \citenamefont {Kwiat},\ and\ \citenamefont
  {Zeilinger}}]{DC_obstacle}%
  \BibitemOpen
  \bibfield  {author} {\bibinfo {author} {\bibfnamefont {K.}~\bibnamefont
  {Mattle}}, \bibinfo {author} {\bibfnamefont {H.}~\bibnamefont {Weinfurter}},
  \bibinfo {author} {\bibfnamefont {P.~G.}\ \bibnamefont {Kwiat}}, \ and\
  \bibinfo {author} {\bibfnamefont {A.}~\bibnamefont {Zeilinger}},\ }\href
  {\doibase 10.1103/PhysRevLett.76.4656} {\bibfield  {journal} {\bibinfo
  {journal} {Phys. Rev. Lett.}\ }\textbf {\bibinfo {volume} {76}},\ \bibinfo
  {pages} {4656} (\bibinfo {year} {1996})}\BibitemShut {NoStop}%
\bibitem [{\citenamefont {L\"utkenhaus}\ \emph {et~al.}(1999)\citenamefont
  {L\"utkenhaus}, \citenamefont {Calsamiglia},\ and\ \citenamefont
  {Suominen}}]{Lutkenhaus_PRA_1999}%
  \BibitemOpen
  \bibfield  {author} {\bibinfo {author} {\bibfnamefont {N.}~\bibnamefont
  {L\"utkenhaus}}, \bibinfo {author} {\bibfnamefont {J.}~\bibnamefont
  {Calsamiglia}}, \ and\ \bibinfo {author} {\bibfnamefont {K.-A.}\ \bibnamefont
  {Suominen}},\ }\href {\doibase 10.1103/PhysRevA.59.3295} {\bibfield
  {journal} {\bibinfo  {journal} {Phys. Rev. A}\ }\textbf {\bibinfo {volume}
  {59}},\ \bibinfo {pages} {3295} (\bibinfo {year} {1999})}\BibitemShut
  {NoStop}%
\bibitem [{\citenamefont {Leibfried}\ \emph {et~al.}(2003)\citenamefont
  {Leibfried}, \citenamefont {Blatt}, \citenamefont {Monroe},\ and\
  \citenamefont {Wineland}}]{DC_obstacle2}%
  \BibitemOpen
  \bibfield  {author} {\bibinfo {author} {\bibfnamefont {D.}~\bibnamefont
  {Leibfried}}, \bibinfo {author} {\bibfnamefont {R.}~\bibnamefont {Blatt}},
  \bibinfo {author} {\bibfnamefont {C.}~\bibnamefont {Monroe}}, \ and\ \bibinfo
  {author} {\bibfnamefont {D.}~\bibnamefont {Wineland}},\ }\href {\doibase
  10.1103/RevModPhys.75.281} {\bibfield  {journal} {\bibinfo  {journal} {Rev.
  Mod. Phys.}\ }\textbf {\bibinfo {volume} {75}},\ \bibinfo {pages} {281}
  (\bibinfo {year} {2003})}\BibitemShut {NoStop}%
\bibitem [{\citenamefont {Vandersypen}\ and\ \citenamefont
  {Chuang}(2005)}]{DC_obstacle3}%
  \BibitemOpen
  \bibfield  {author} {\bibinfo {author} {\bibfnamefont {L.~M.~K.}\
  \bibnamefont {Vandersypen}}\ and\ \bibinfo {author} {\bibfnamefont {I.~L.}\
  \bibnamefont {Chuang}},\ }\href {\doibase 10.1103/RevModPhys.76.1037}
  {\bibfield  {journal} {\bibinfo  {journal} {Rev. Mod. Phys.}\ }\textbf
  {\bibinfo {volume} {76}},\ \bibinfo {pages} {1037} (\bibinfo {year}
  {2005})}\BibitemShut {NoStop}%
\bibitem [{\citenamefont {Braunstein}\ and\ \citenamefont
  {Kimble}(2000)}]{Braunstein_PRA_2000}%
  \BibitemOpen
  \bibfield  {author} {\bibinfo {author} {\bibfnamefont {S.~L.}\ \bibnamefont
  {Braunstein}}\ and\ \bibinfo {author} {\bibfnamefont {H.~J.}\ \bibnamefont
  {Kimble}},\ }\href {\doibase 10.1103/PhysRevA.61.042302} {\bibfield
  {journal} {\bibinfo  {journal} {Phys. Rev. A}\ }\textbf {\bibinfo {volume}
  {61}},\ \bibinfo {pages} {042302} (\bibinfo {year} {2000})}\BibitemShut
  {NoStop}%
\bibitem [{\citenamefont {Einstein}\ \emph {et~al.}(1935)\citenamefont
  {Einstein}, \citenamefont {Podolsky},\ and\ \citenamefont
  {Rosen}}]{Einstein_PRA_1935}%
  \BibitemOpen
  \bibfield  {author} {\bibinfo {author} {\bibfnamefont {A.}~\bibnamefont
  {Einstein}}, \bibinfo {author} {\bibfnamefont {B.}~\bibnamefont {Podolsky}},
  \ and\ \bibinfo {author} {\bibfnamefont {N.}~\bibnamefont {Rosen}},\ }\href
  {\doibase 10.1103/PhysRev.47.777} {\bibfield  {journal} {\bibinfo  {journal}
  {Phys. Rev.}\ }\textbf {\bibinfo {volume} {47}},\ \bibinfo {pages} {777}
  (\bibinfo {year} {1935})}\BibitemShut {NoStop}%
\bibitem [{\citenamefont {Hao}\ \emph {et~al.}(2021)\citenamefont {Hao},
  \citenamefont {Shi}, \citenamefont {Li}, \citenamefont {Shapiro},
  \citenamefont {Zhuang},\ and\ \citenamefont {Zhang}}]{Hao_PRL_2021}%
  \BibitemOpen
  \bibfield  {author} {\bibinfo {author} {\bibfnamefont {S.}~\bibnamefont
  {Hao}}, \bibinfo {author} {\bibfnamefont {H.}~\bibnamefont {Shi}}, \bibinfo
  {author} {\bibfnamefont {W.}~\bibnamefont {Li}}, \bibinfo {author}
  {\bibfnamefont {J.~H.}\ \bibnamefont {Shapiro}}, \bibinfo {author}
  {\bibfnamefont {Q.}~\bibnamefont {Zhuang}}, \ and\ \bibinfo {author}
  {\bibfnamefont {Z.}~\bibnamefont {Zhang}},\ }\href {\doibase
  10.1103/PhysRevLett.126.250501} {\bibfield  {journal} {\bibinfo  {journal}
  {Phys. Rev. Lett.}\ }\textbf {\bibinfo {volume} {126}},\ \bibinfo {pages}
  {250501} (\bibinfo {year} {2021})}\BibitemShut {NoStop}%
\bibitem [{\citenamefont {Barzanjeh}\ \emph {et~al.}(2013)\citenamefont
  {Barzanjeh}, \citenamefont {Pirandola},\ and\ \citenamefont
  {Weedbrook}}]{Barzanjeh_PRA_2013}%
  \BibitemOpen
  \bibfield  {author} {\bibinfo {author} {\bibfnamefont {S.}~\bibnamefont
  {Barzanjeh}}, \bibinfo {author} {\bibfnamefont {S.}~\bibnamefont
  {Pirandola}}, \ and\ \bibinfo {author} {\bibfnamefont {C.}~\bibnamefont
  {Weedbrook}},\ }\href {\doibase 10.1103/PhysRevA.88.042331} {\bibfield
  {journal} {\bibinfo  {journal} {Phys. Rev. A}\ }\textbf {\bibinfo {volume}
  {88}},\ \bibinfo {pages} {042331} (\bibinfo {year} {2013})}\BibitemShut
  {NoStop}%
\bibitem [{\citenamefont {Adesso}\ \emph {et~al.}(2014)\citenamefont {Adesso},
  \citenamefont {Ragy},\ and\ \citenamefont {Lee}}]{Adesso_OSID_2014}%
  \BibitemOpen
  \bibfield  {author} {\bibinfo {author} {\bibfnamefont {G.}~\bibnamefont
  {Adesso}}, \bibinfo {author} {\bibfnamefont {S.}~\bibnamefont {Ragy}}, \ and\
  \bibinfo {author} {\bibfnamefont {A.~R.}\ \bibnamefont {Lee}},\ }\href
  {\doibase 10.1142/s1230161214400010} {\bibfield  {journal} {\bibinfo
  {journal} {Open Systems \& Information Dynamics}\ }\textbf {\bibinfo {volume}
  {21}},\ \bibinfo {pages} {1440001} (\bibinfo {year} {2014})}\BibitemShut
  {NoStop}%
\bibitem [{\citenamefont {Kim}\ \emph {et~al.}(2002)\citenamefont {Kim},
  \citenamefont {Lee},\ and\ \citenamefont {Munro}}]{Kim_PRA_2002}%
  \BibitemOpen
  \bibfield  {author} {\bibinfo {author} {\bibfnamefont {M.~S.}\ \bibnamefont
  {Kim}}, \bibinfo {author} {\bibfnamefont {J.}~\bibnamefont {Lee}}, \ and\
  \bibinfo {author} {\bibfnamefont {W.~J.}\ \bibnamefont {Munro}},\ }\href
  {\doibase 10.1103/PhysRevA.66.030301} {\bibfield  {journal} {\bibinfo
  {journal} {Phys. Rev. A}\ }\textbf {\bibinfo {volume} {66}},\ \bibinfo
  {pages} {030301} (\bibinfo {year} {2002})}\BibitemShut {NoStop}%
\bibitem [{\citenamefont {Wang}\ \emph {et~al.}(2020)\citenamefont {Wang},
  \citenamefont {Pi}, \citenamefont {Huang}, \citenamefont {Li}, \citenamefont
  {Shao}, \citenamefont {Yang}, \citenamefont {Liu}, \citenamefont {Zhang},
  \citenamefont {Zhang},\ and\ \citenamefont {Xu}}]{Wang_OptEx_2020}%
  \BibitemOpen
  \bibfield  {author} {\bibinfo {author} {\bibfnamefont {H.}~\bibnamefont
  {Wang}}, \bibinfo {author} {\bibfnamefont {Y.}~\bibnamefont {Pi}}, \bibinfo
  {author} {\bibfnamefont {W.}~\bibnamefont {Huang}}, \bibinfo {author}
  {\bibfnamefont {Y.}~\bibnamefont {Li}}, \bibinfo {author} {\bibfnamefont
  {Y.}~\bibnamefont {Shao}}, \bibinfo {author} {\bibfnamefont {J.}~\bibnamefont
  {Yang}}, \bibinfo {author} {\bibfnamefont {J.}~\bibnamefont {Liu}}, \bibinfo
  {author} {\bibfnamefont {C.}~\bibnamefont {Zhang}}, \bibinfo {author}
  {\bibfnamefont {Y.}~\bibnamefont {Zhang}}, \ and\ \bibinfo {author}
  {\bibfnamefont {B.}~\bibnamefont {Xu}},\ }\href {\doibase 10.1364/OE.404611}
  {\bibfield  {journal} {\bibinfo  {journal} {Optics express}\ }\textbf
  {\bibinfo {volume} {28}},\ \bibinfo {pages} {32882–32893} (\bibinfo {year}
  {2020})}\BibitemShut {NoStop}%
\bibitem [{\citenamefont {Ralph}\ and\ \citenamefont
  {Huntington}(2002)}]{Ralph_PRA_2002}%
  \BibitemOpen
  \bibfield  {author} {\bibinfo {author} {\bibfnamefont {T.~C.}\ \bibnamefont
  {Ralph}}\ and\ \bibinfo {author} {\bibfnamefont {E.~H.}\ \bibnamefont
  {Huntington}},\ }\href {\doibase 10.1103/PhysRevA.66.042321} {\bibfield
  {journal} {\bibinfo  {journal} {Phys. Rev. A}\ }\textbf {\bibinfo {volume}
  {66}},\ \bibinfo {pages} {042321} (\bibinfo {year} {2002})}\BibitemShut
  {NoStop}%
\bibitem [{\citenamefont {Jing}\ \emph {et~al.}(2003)\citenamefont {Jing},
  \citenamefont {Zhang}, \citenamefont {Yan}, \citenamefont {Zhao},
  \citenamefont {Xie},\ and\ \citenamefont {Peng}}]{Jing_PRL_2003}%
  \BibitemOpen
  \bibfield  {author} {\bibinfo {author} {\bibfnamefont {J.}~\bibnamefont
  {Jing}}, \bibinfo {author} {\bibfnamefont {J.}~\bibnamefont {Zhang}},
  \bibinfo {author} {\bibfnamefont {Y.}~\bibnamefont {Yan}}, \bibinfo {author}
  {\bibfnamefont {F.}~\bibnamefont {Zhao}}, \bibinfo {author} {\bibfnamefont
  {C.}~\bibnamefont {Xie}}, \ and\ \bibinfo {author} {\bibfnamefont
  {K.}~\bibnamefont {Peng}},\ }\href {\doibase 10.1103/PhysRevLett.90.167903}
  {\bibfield  {journal} {\bibinfo  {journal} {Phys. Rev. Lett.}\ }\textbf
  {\bibinfo {volume} {90}},\ \bibinfo {pages} {167903} (\bibinfo {year}
  {2003})}\BibitemShut {NoStop}%
\bibitem [{\citenamefont {Barnett}\ and\ \citenamefont
  {Radmore}(2002)}]{Barnett_book_2002}%
  \BibitemOpen
  \bibfield  {author} {\bibinfo {author} {\bibfnamefont {S.}~\bibnamefont
  {Barnett}}\ and\ \bibinfo {author} {\bibfnamefont {P.}~\bibnamefont
  {Radmore}},\ }\href {\doibase 10.1093/acprof:oso/9780198563617.001.0001}
  {\emph {\bibinfo {title} {Methods in Theoretical Quantum Optics}}}\ (\bibinfo
   {publisher} {Oxford Scholarship Online},\ \bibinfo {year}
  {2002})\BibitemShut {NoStop}%
\bibitem [{\citenamefont {Wigner}(1932)}]{Wigner_PR_1932}%
  \BibitemOpen
  \bibfield  {author} {\bibinfo {author} {\bibfnamefont {E.}~\bibnamefont
  {Wigner}},\ }\href {\doibase 10.1103/PhysRev.40.749} {\bibfield  {journal}
  {\bibinfo  {journal} {Phys. Rev.}\ }\textbf {\bibinfo {volume} {40}},\
  \bibinfo {pages} {749} (\bibinfo {year} {1932})}\BibitemShut {NoStop}%
\bibitem [{\citenamefont {Holevo}(1998)}]{Holevo-1998}%
  \BibitemOpen
  \bibfield  {author} {\bibinfo {author} {\bibfnamefont {A.}~\bibnamefont
  {Holevo}},\ }\href {\doibase 10.1109/18.651037} {\bibfield  {journal}
  {\bibinfo  {journal} {IEEE Transactions on Information Theory}\ }\textbf
  {\bibinfo {volume} {44}},\ \bibinfo {pages} {269} (\bibinfo {year}
  {1998})}\BibitemShut {NoStop}%
\bibitem [{\citenamefont {Caves}\ and\ \citenamefont
  {Drummond}(1994)}]{drummond_Caves-1994}%
  \BibitemOpen
  \bibfield  {author} {\bibinfo {author} {\bibfnamefont {C.~M.}\ \bibnamefont
  {Caves}}\ and\ \bibinfo {author} {\bibfnamefont {P.~D.}\ \bibnamefont
  {Drummond}},\ }\href {\doibase 10.1103/RevModPhys.66.481} {\bibfield
  {journal} {\bibinfo  {journal} {Rev. Mod. Phys.}\ }\textbf {\bibinfo {volume}
  {66}},\ \bibinfo {pages} {481} (\bibinfo {year} {1994})}\BibitemShut
  {NoStop}%
\bibitem [{\citenamefont {Yuen}\ and\ \citenamefont
  {Ozawa}(1993)}]{Yuen_Ozawa-1993}%
  \BibitemOpen
  \bibfield  {author} {\bibinfo {author} {\bibfnamefont {H.~P.}\ \bibnamefont
  {Yuen}}\ and\ \bibinfo {author} {\bibfnamefont {M.}~\bibnamefont {Ozawa}},\
  }\href {\doibase 10.1103/PhysRevLett.70.363} {\bibfield  {journal} {\bibinfo
  {journal} {Phys. Rev. Lett.}\ }\textbf {\bibinfo {volume} {70}},\ \bibinfo
  {pages} {363} (\bibinfo {year} {1993})}\BibitemShut {NoStop}%
\bibitem [{\citenamefont {van Loock}\ and\ \citenamefont
  {Braunstein}(2000)}]{Braunstein_PRL_2000}%
  \BibitemOpen
  \bibfield  {author} {\bibinfo {author} {\bibfnamefont {P.}~\bibnamefont {van
  Loock}}\ and\ \bibinfo {author} {\bibfnamefont {S.~L.}\ \bibnamefont
  {Braunstein}},\ }\href {\doibase 10.1103/PhysRevLett.84.3482} {\bibfield
  {journal} {\bibinfo  {journal} {Phys. Rev. Lett.}\ }\textbf {\bibinfo
  {volume} {84}},\ \bibinfo {pages} {3482} (\bibinfo {year}
  {2000})}\BibitemShut {NoStop}%
\bibitem [{\citenamefont {Adesso}\ \emph {et~al.}(2007)\citenamefont {Adesso},
  \citenamefont {Serafini},\ and\ \citenamefont
  {Illuminati}}]{Adesso_arXiv_2006}%
  \BibitemOpen
  \bibfield  {author} {\bibinfo {author} {\bibfnamefont {G.}~\bibnamefont
  {Adesso}}, \bibinfo {author} {\bibfnamefont {A.}~\bibnamefont {Serafini}}, \
  and\ \bibinfo {author} {\bibfnamefont {F.}~\bibnamefont {Illuminati}},\
  }\href {\doibase 10.1088/1367-2630/9/3/060} {\bibfield  {journal} {\bibinfo
  {journal} {New Journal of Physics}\ }\textbf {\bibinfo {volume} {9}},\
  \bibinfo {pages} {60} (\bibinfo {year} {2007})}\BibitemShut {NoStop}%
\bibitem [{\citenamefont {Adesso}\ and\ \citenamefont
  {Illuminati}(2007)}]{Adesso_JPA_2007}%
  \BibitemOpen
  \bibfield  {author} {\bibinfo {author} {\bibfnamefont {A.}~\bibnamefont
  {Adesso}}\ and\ \bibinfo {author} {\bibfnamefont {F.}~\bibnamefont
  {Illuminati}},\ }\href {\doibase 10.1088/1751-8113/40/28/s01} {\bibfield
  {journal} {\bibinfo  {journal} {Journal of Physics A: Mathematical and
  Theoretical}\ }\textbf {\bibinfo {volume} {40}},\ \bibinfo {pages} {7821}
  (\bibinfo {year} {2007})}\BibitemShut {NoStop}%
\end{thebibliography}%

\end{document}